\documentclass[fleqn,usenatbib]{mnras}
\usepackage[T1]{fontenc}

\DeclareRobustCommand{\VAN}[3]{#2}
\let\VANthebibliography\thebibliography
\def\thebibliography{\DeclareRobustCommand{\VAN}[3]{##3}\VANthebibliography}

\usepackage{graphicx}	
\usepackage{amsmath}	
\usepackage{amssymb}	
\usepackage{bm}         
\usepackage{tikz}

\newcommand{\lya}{\mbox{Lyman-$\alpha$}}
\newcommand{\hi}{\mbox{H~\textsc{i}}}
\newcommand{\hii}{\mbox{H~\textsc{ii}}}
\newcommand{\hei}{\mbox{He~\textsc{i}}}
\newcommand{\heii}{\mbox{He~\textsc{ii}}}
\newcommand{\heiii}{\mbox{He~\textsc{iii}}}

\newcommand{\at}{\textsc{aton}}

\usepackage[normalem]{ulem}
\newcommand{\atmf}{\textsc{aton-he}}

\newcommand{\sherr}{\textsc{sherwood-relics}}

\definecolor{notecolor}{rgb}{0.8,0,0}

\usepackage{newtxtext,newtxmath}

\newcommand{\citeyearname}[1]{\citeauthor{#1} \citeyear{#1}}
\title{The impact of faint AGN discovered by JWST on reionization}
\author[]{
Shikhar Asthana$^{1}$\thanks{E-mail: sa2001@cam.ac.uk},
Martin G. Haehnelt$^{1}$,  
Girish Kulkarni$^{2}$,
James S. Bolton$^{3}$,  \newauthor
Prakash Gaikwad$^{4}$, 
Laura C. Keating$^{5}$
and Ewald Puchwein$^{6}$
\\
$^{1}$Kavli Institute for Cosmology and Institute of Astronomy, Madingley Road, Cambridge, CB3 0HA, UK\\
$^{2}$Tata Institute of Fundamental Research, Homi Bhabha Road, Mumbai 400005, India\\
$^{3}$School of Physics and Astronomy, University of Nottingham, University Park, Nottingham, NG7 2RD, UK\\
$^{4}$ Max-Planck-Institut für Astronomie, Königstuhl 17, D-69117 Heidelberg, Germany\\
$^{5}$Institute for Astronomy, University of Edinburgh, Blackford Hill, Edinburgh, EH9 3HJ, UK\\
$^{6}$Leibniz-Institut f\"ur Astrophysik Potsdam, An der Sternwarte 16, 14482 Potsdam, Germany\\
}
\date{Accepted ---. Received ---; in original form ---}
\pubyear{2024}

\begin{document}
\label{firstpage}
\pagerange{\pageref{firstpage}--\pageref{lastpage}}
\maketitle

\begin{abstract}
The relative contribution of emission from stellar sources and accretion onto supermassive black holes to reionization has been brought into focus again by the apparent high abundance of faint Active Galactic Nuclei (AGN) at $4\lesssim z\lesssim11$ uncovered by JWST. We investigate here the contribution of these faint AGN to hydrogen and the early stages of helium reionization using the GPU-based radiative transfer code \atmf\ by post-processing a cosmological hydrodynamical simulation from the \sherr\ suite of simulations. We study four models: two galaxy-only late-end reionization models, a QSO-assisted and a QSO-only model. In the QSO-assisted model, 1\% of the haloes host AGN, with AGN luminosities scaled to contribute 17\% of the total hydrogen-ionizing emissivity. In the QSO-only model, quasars account for all the hydrogen-ionizing emissivity, with 10\% of the haloes hosting AGN. The SED of AGN is assumed to be a power-law with $\alpha=-1.7$ each with a 10~Myr lifetime. All models are calibrated to the observed mean \lya\ forest transmission at $5\lesssim z\lesssim6.2$. The QSO-assisted model requires an emissivity similar to the galaxy-only models and fits the observed distribution of the \lya\ optical depths well. The QSO-only model is inconsistent with the observed \lya\ optical depths distribution, and produces excessively high IGM temperatures at $z\lesssim 5$ due to an early onset of \heii\ reionization, unless the escape fraction of \heii-ionizing photons is assumed to be low. Our results suggest that a modest AGN contribution to reionization aligns with the \lya\ forest data, whereas an AGN dominated scenario is difficult to reconcile.
\end{abstract}


\section{Introduction}

The first sources that emitted sufficiently energetic UV photons reionized hydrogen, allowing the Universe to become transparent to UV photons. This process of reionization started around individual UV sources \citep{Arons-1970, Ciardi2005, loeb-2013, Mcquinn2016, Dayal2018}, creating ionized bubbles which eventually coalesced, leading to an almost completely ionized intergalactic medium (IGM). However, what process produced the ionizing UV emissions is still an open question. The most plausible candidates are massive stars and accretion onto supermassive black holes. 

A lively discussion of this question goes back to the late 1980s when neither the space density of galaxies nor that of Active Galactic Nuclei (AGN) at the relevant redshifts had yet been established \citep{Shapiro1987, SHAPIRO1989, Escude-1990, Songaila1990, MEIKSIN1991, Madau1991}. Additionally, the fraction of ionizing photons produced in massive stars escaping from galaxies was suspected to be low and was then and is now still very uncertain. Furthermore, the fraction of hydrogen ionizing photons escaping from quasi-stellar objects (QSOs) is known to approach unity, at least if they are bright \citep{Cristiani2016, Grazian2018, Romano2019}.  
 
Traditionally, the search for high-redshift QSOs has been led by wide-field surveys, such as SDSS and CFHQS, using ground-based telescopes \citep{2000AJ....120.1167F, 2007AJ....134.2435W, 2010AJ....139..906W, 2016ApJ...833..222J}.  In the last decade, newer surveys have added to the sample and extended the redshift reach. This includes PS1 \citep{2016ApJS..227...11B, 2017NatSR...741617K, 2017ApJ...849...91M, 2017MNRAS.466.4568T}, SHELLQs \citep{2016ApJ...828...26M, 2018PASJ...70S..35M, 2018ApJS..237....5M}, DES \citep{2015MNRAS.454.3952R, 2017MNRAS.468.4702R, 2019MNRAS.487.1874R} and DELS \citep{2017ApJ...839...27W, 2019ApJ...884...30W}.  Recently, a fruitful approach to high-redshift quasar discovery has been to use a combination of optical and mid-IR surveys. Examples of this include UKIDSS \citep{2011Natur.474..616M}, UHS \citep{2017ApJ...839...27W}, VIKINGS \citep{2013ApJ...779...24V}, VST ATLAS \citep{2015MNRAS.451L..16C, 2018MNRAS.478.1649C} and VHS \citep{2019MNRAS.484.5142P}.  There has been occasional use of other wavelengths, such as X-rays \citep{2015A&A...578A..83G} or radio \citep{2006ApJ...652..157M}. The highest-redshift QSO known by this method is at $z = 7.64$ with $M_{1450}=-26.13$, discovered by combining data from the PS1, DELS, VHS, and WISE surveys \citep{2021ApJ...907L...1W}.  The overall conclusion of these surveys is that the number density of bright, high-redshift QSOs drops with increasing redshift to the extent that their contribution to reionization is sub-dominant compared to star-forming galaxies \citep{2018ApJ...869..150M, 2019MNRAS.488.1035K}.

Nevertheless, the idea of a negligible contribution of QSOs to reionization critically depends on the number density of faint AGN, with UV magnitude as low as $M_{1450}<-18$. Ground-based surveys have largely only probed the bright end of the AGN luminosity function. Consequently, most conclusions in the literature about the contribution of AGN to reionization hinge on an uncertain extrapolation of the luminosity function to the faint end \citep{2019MNRAS.488.1035K}.  The suggestion that the faint-end of the AGN luminosity function at high redshifts might be steeper than what is indicated by optical and mid-IR surveys first came from \citet{2015A&A...578A..83G}, who used X-ray measurements to discover faint AGN. In just over two years of operation, JWST has given fresh support to this idea by finding many faint AGN, primarily thanks to the sensitivity and high resolution of the NIRCam and NIRSpec instruments.

JWST has yielded several hundreds of AGN with $M_{1450}\sim -20$ or even fainter \citep{2023ApJ...942L..17O, 2023ApJ...959...39H, 2023ApJ...952..142F, 2023arXiv230801230M, 2023arXiv230607320L, 2023ApJ...957L...7K, 2023arXiv230811609F, 2023ApJ...955L..24G, 2024ApJ...964...39G, 2024ApJ...963..129M, 2024arXiv240610341A}.  Indeed, the two spectroscopically confirmed highest-redshift AGN now known are an object with UV magnitude of about $-22$, discovered using JWST at $z=8.679$ \citep{2023ApJ...953L..29L} and a lensed X-ray luminous AGN at $z=10.1$, discovered using the Chandra X-ray Observatory and spectroscopically confirmed by JWST \citep{2023ApJ...955L..24G, 2024NatAs...8..126B}. Several, although not all, of these high-redshift AGN are ``Little Red Dots'' \citep[LRDs;][]{2024arXiv240610341A,2024ApJ...963..129M}.  LRDs have compact, quasi-stellar morphology, with red colours at 2--5$\;\mu m$ but often rather blue colours in the UV \citep{2024arXiv240610341A}. Whether or not LRDs are AGN is still debated \citep{Kokubo2024, Baggen2024}, although 80\% of the 50 or so LRDs that have been spectroscopically observed show evidence of AGN-like broad emission lines \citep{2024arXiv240302304W, 2023ApJ...954L...4K, 2024ApJ...964...39G, 2023ApJ...952..142F, 2023arXiv231203065K, 2024arXiv240403576K}. As a result of these spectacular discoveries, a scenario in which the faint end of the AGN luminosity function is steep at high redshifts is now gaining traction \citep{2023arXiv230801230M, 2024arXiv240610341A, Grazian2024, Madau2024}. The uncertainty on the faint-end of the AGN UV luminosity function continues to be large \citep{2023arXiv230801230M}, and it is unclear if the escape fraction of hydrogen-ionizing photons from AGN is as high as from bright QSOs \citep{2017MNRAS.465..302M, 2020ApJ...897...41S}. Nonetheless, the possibility of a large number density of AGN at high redshifts makes it imperative to examine their contribution to reionization and their consistency with measurements of the ionization and thermal state of the IGM. Indeed, the discovery of these faint high-redshift AGN has led to a resurgence of interest in the possibility that hydrogen reionization is solely driven by QSOs \citep{Madau2024}, following many earlier such suggestions \citep{Giallongo1994, Meiksin-1993, Zoltan1998, Madau2015}. 

As discussed in detail by \citet{Madau2024}, models where hydrogen reionization is driven solely by AGN/QSOs have to avoid reionizing too much \heii\ too early to be consistent with the measurements of the thermal history of the IGM as well as the evolution of the \heii\ \lya\ opacity, which suggest that \heii\ reionization is completed in the redshift range $2.7<z<4$ \citep{Worseck2016, Laplante2018, Puchwein2019, Gaikwad2021, Becker2021, Makan2021, Makan2022, Basu2024}.


Recently, improved \lya\ forest data has  provided critical quantitative constraints on both the timing of late stages of reionization as well as on a possible QSO\footnote{In the text, we are using the term (faint) AGN to denote sources with  QSO-like  spectra to account for the fact that the sources we have included are fainter than traditional QSOs.} contribution \citep{Chardin2017, DAloisio2017, bosman2018, Bosman-2022, Kulkarni-2019, Keating-2019, Gaikwad-2020, Gaikwad-2023, Zhu-2022, Zhu-2023, Zhu2024, DOdorico-2023}. These models have demonstrated that AGN-dominated models struggle to reconcile with the thermal history of the IGM with a typical QSO SED. In this work, we provide a significant extension to this research by considering AGN-assisted models, incorporating new observational and numerical advancements and making, in particular, a quantitative comparison to the latest measurements of the Lyman-$\alpha$ flux distribution. A key motivation for our analysis was thereby to investigate the impact of the recently discovered population of faint AGN identified by JWST. The role of these faint sources in reionization remains unexplored, and our work provides the first radiative transfer simulations that explicitly account for their contribution. To do so, we post-processed a simulation from the \sherr\ suite of simulations \citep{Puchwein-2023} with the GPU-based radiative transfer code \atmf\ \citep{Asthana2024} to explore possible effects of the population of faint AGN uncovered by JWST on reionization.

We create two models; one with a 17\%  contribution to hydrogen reionization by faint AGN, while in the other, hydrogen reionization is solely driven by faint AGN. We compare these QSO-assisted and QSO-only models to two of our previously studied galaxy-only models from \citet{Asthana2024}---the fiducial model of that work and the `Oligarchic' model with a relatively high mass cutoff for the haloes hosting ionizing sources. All simulations have been calibrated to the observed mean \lya\ forest transmission at  $5\lesssim z\lesssim6.2$. 

The paper is structured as follows. In Section~\ref{sec:simulation_suite}, we describe the simulations used in this work, the calibration of these simulations to the observed mean \lya\ forest transmission, and our modelling of QSOs. Section~\ref{sec:QSOassist} describes the QSO-assisted and QSO-only reionization model results. In Section~\ref{sec:discussion}, we discuss the effect of AGN on ionized \hii\ regions, compare the Oligarchic model to the QSO-assisted models, comment on the \heiii\ ionization, the UV luminosity function of galaxies and QSOs, and the possible black hole masses of the AGN in our models. The paper concludes in Section~\ref{sec:conclusion}. Throughout this work, we assume a $\Lambda$CDM cosmology with parameter values from \citet{planck-2014} ($\Omega_m=0.308, \Omega_\Lambda=0.6982, h=0.678, \Omega_b=0.482, \sigma_8=0.829$, and $n=0.961$).

\section{Simulation setup} \label{sec:simulation_suite}

Our models are created by post-processing cosmological hydrodynamical simulations taken from the \mbox{\sherr}\ \citep{Puchwein-2023} suite of simulations. This is done using the GPU-based M1-closure radiative transfer code \atmf\, details of which can be found in \citet{Asthana2024}. We will discuss the salient features of the simulations below (see \citeyearname{Gnedin2022} for a general review of modelling reionization). \begin{figure*}
  \centering
  \includegraphics[width=\columnwidth]{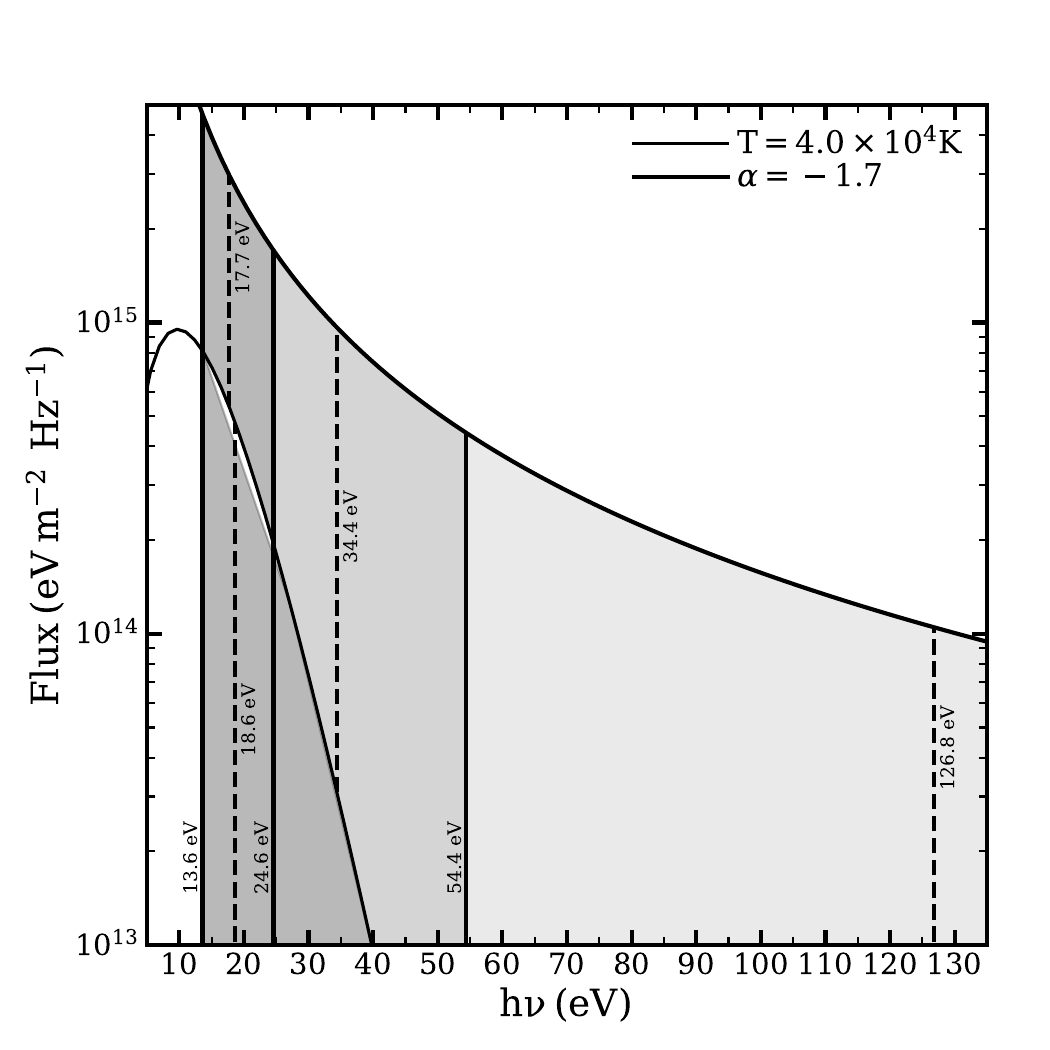}
  \includegraphics[width=\columnwidth]{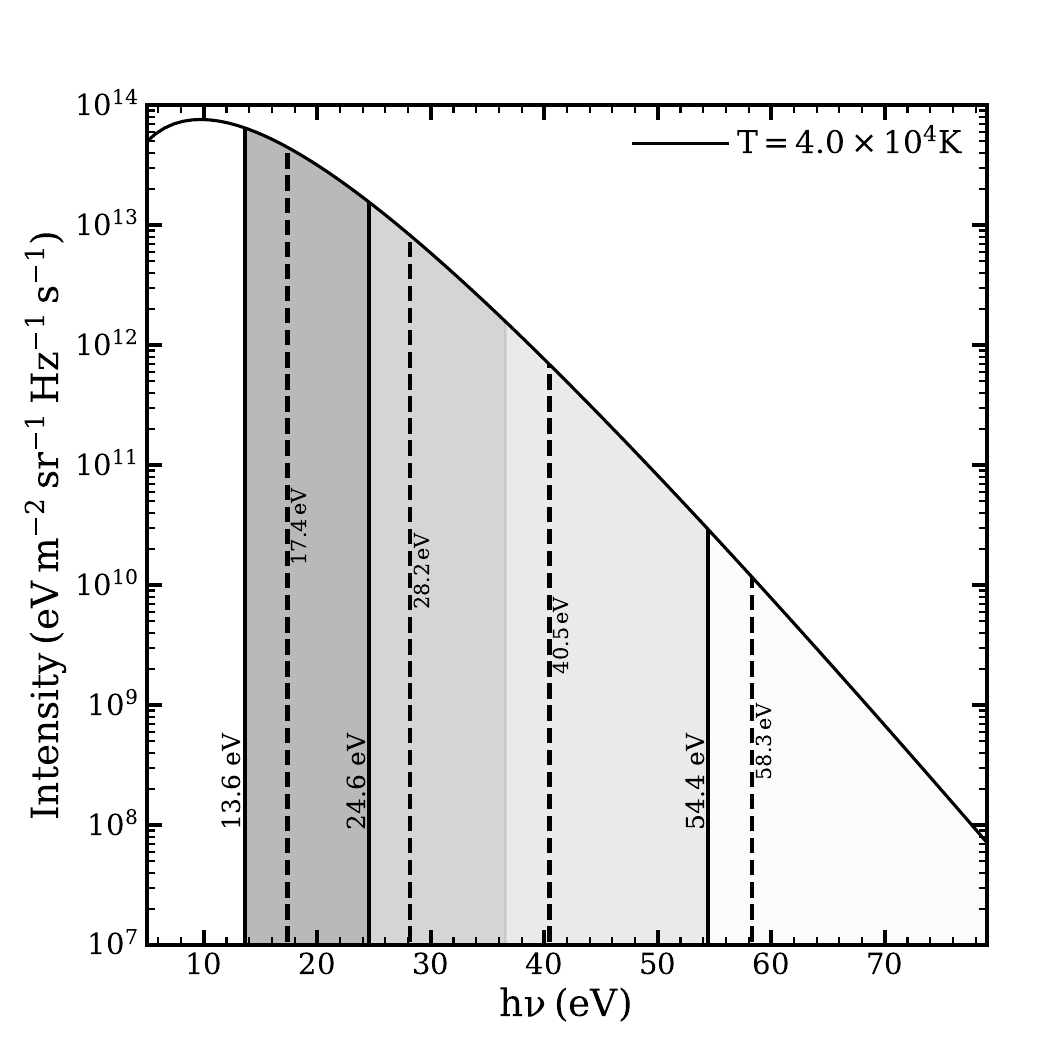}
 \caption{\textit{Left:} the assumed black-body spectrum for galaxies and power-law spectrum for the QSO-assisted and QSO-only models in this work. The galaxy spectrum is mono-frequency. \textit{Right:} the black-body spectrum for the galaxies-only simulations, i.e. the fiducial and Oligarchic models. The galaxy spectrum is divided into four frequency bins. The temperature of the black-body spectrum for the galaxies is $T=4\times10^4$~K. The power law for AGN is $f_\nu \propto \nu^\alpha$, with $\alpha=-1.7$ \citep{Lusso2015}, and an arbitrary normalization in this figure. The frequency bins chosen are shown as the shaded regions, each having a distinct colour, with the highest energy bin extending to infinity. Dashed lines within each shaded region represent the average photon energy of the photon for the corresponding bin. The ionization energies for \hi\ (13.6~eV), \hei\ (24.6~eV), and \heii\ (54.4~eV) are marked by solid vertical lines. 
 }
  \label{fig:BB-spec}
\end{figure*}
\subsection{Cosmological hydrodynamical simulations}
The \sherr\ suite  of simulations\citep{Puchwein-2023} were run using the Tree-PM SPH code \mbox{\textsc{p-gadget-3}} \citep[][an extended version of  \textsc{p-gadget-2}]{2005MNRAS.364.1105S} . The suite contains models with varying box sizes, and we are focusing here on the $160$~cMpc$/h$ box, which strikes the best balance between resolution and volume for comparisons with \lya\ forest data. The model includes $2\times2048^3$ gas and dark matter particles. Starting at a redshift of $z=99$, snapshots are saved every $40$~Myr down to redshift $z=4$. Star formation is implemented using a simplified recipe (invoked using the \texttt{QUICK\_LYALPHA} compile-time flag in \textsc{p-gadget-3}), where gas particles exceeding a density threshold of $\Delta = 10^3$, with a temperature $\lesssim 10^5$~K, are removed from the hydrodynamic calculations and are converted into star particles \citep{Viel-2004}.  A uniform UV background is integrated into the simulations as described by \citet{Puchwein2019}. For the post-processing, we project the gas density onto a grid of $2048^3$ cells, as \atmf\ requires a uniform Cartesian grid. 

\subsection{Radiative transfer with ATON-HE}
\atmf\ is a GPU-based radiative transfer code that takes angular moments of the cosmological radiative transfer equation, thus reducing the dimensionality \citep{Aubert-2008, Aubert-2010, Asthana2024}. The equations are truncated at second order and are closed using the M1 relation \citep{Levermore-1984}, converting the RT equation into a set of equations for energy and flux. The code follows the ionization states of hydrogen and helium. In the simulation, after establishing the matter distribution, sources are placed at the location of the dark matter haloes following a source model. The combined volume ionizing emissivity of the sources in the simulation is a free parameter modulated to fit the mean \lya\ forest transmission between $5\lesssim z \lesssim 6.2$ \citep{Bosman-2022}. These measurements have been taken from high-SNR, high-resolution quasar spectra from the E-XQR-30 data set \citep{DOdorico-2023}. A multi-frequency simulation including helium with ionizing sources emitting from $z=19.3$ to $z=4.9$ requires approximately $900$~GPU-hours on $32$ NVIDIA A100 GPUs. 
\subsection{Implementation of AGN as ionizing sources in ATON-HE}\label{sec:quasar}

For the ionizing spectrum of the stellar sources in \atmf, we have assumed a black-body spectrum. For the QSO models discussed here, we have modelled the galaxy spectrum with a single frequency bin as seen in the left panel of Figure~\ref{fig:BB-spec}. This was done to save computational resources (see \citeyearname{Asthana2024} for a discussion of mono-frequency versus multi-frequency simulations of stellar sources for hydrogen reionization). The galaxy-only simulations have four frequency bins, as shown in the right panel of Figure~\ref{fig:BB-spec}, and are the fiducial and Oligarchic models from \citet{Asthana2024}. To incorporate faint AGN in our simulations, we randomly choose a small fraction of haloes to host the AGN. Bluewards of the Lyman continuum edge at 912 \AA, we model the SED of the AGN as commonly assumed as a power law spectrum with spectral index $\alpha=-1.7$ \citep{Lusso2015} (See \citealt{Madau2024} for a discussion of the observational uncertainty in that part of the  SED of AGN.) 
The AGN spectra are divided into three bins, as shown in Figure~\ref{fig:BB-spec}. The galaxy and AGN spectra are divided into four independent frequency bins. The ionizing photons in the simulations have thus four energies: 18.6~eV, 17.7~eV, 34.4~eV, and 126.8~eV. Dividing the power-law spectra of AGN into three bins is the minimum necessary to track hydrogen and helium ionization states. 
\begin{figure*}
  \centering
  \includegraphics[width=1.5\columnwidth,,trim={0.5cm 0.5cm 0cm 0cm},clip]{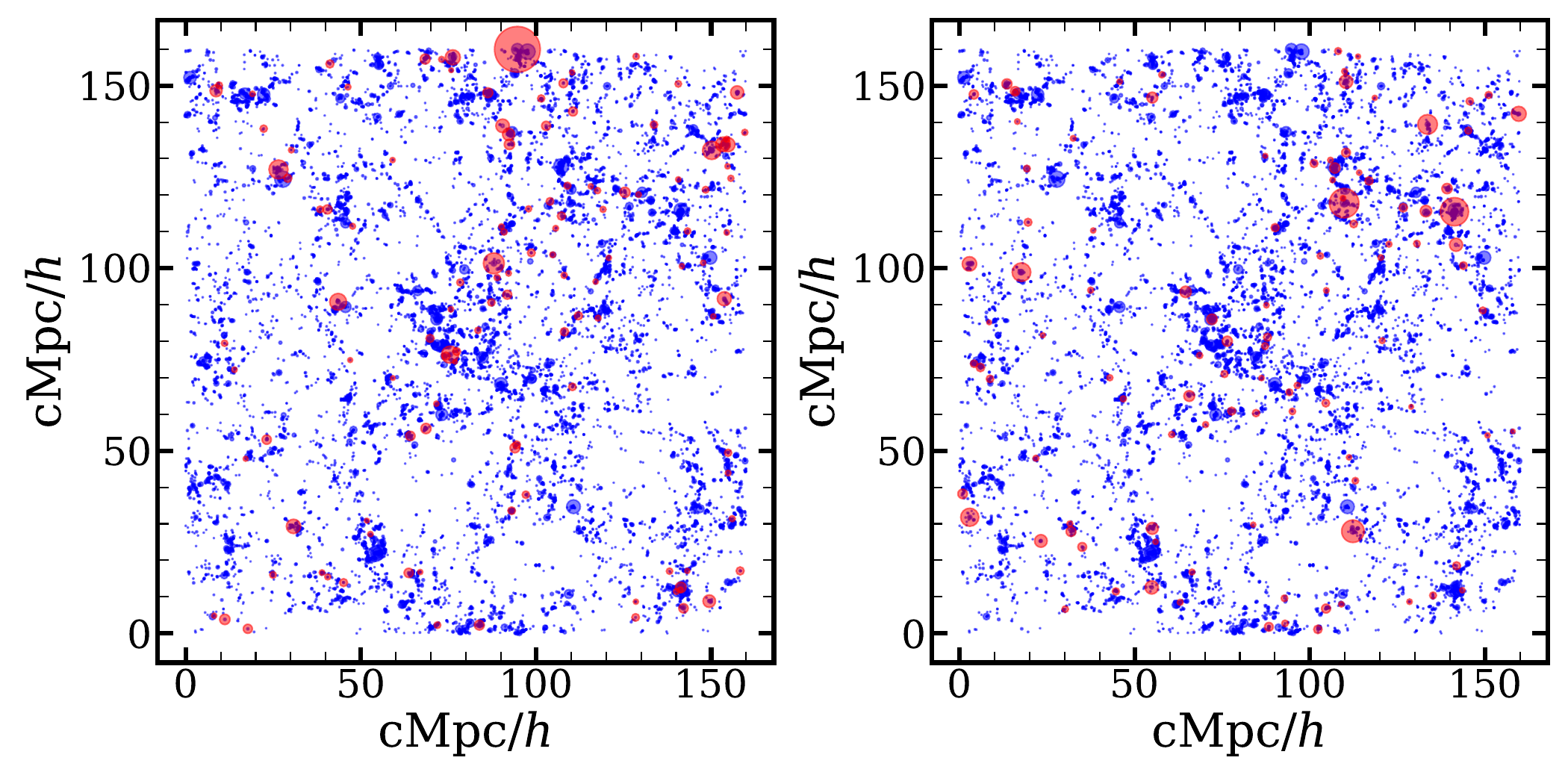}
  \put(-200,160){\makebox(0,0){\colorbox{white}{\textbf{z=6.10}}}} 
  \put(-20,160){\makebox(0,0){\colorbox{white}{\textbf{z=6.05}}}} 
 \caption{ The spatial distribution of galaxies (blue) and AGN (red) in our QSO-assisted model at redshifts $z=6.10$ (left panel) and $z=6.05$ (right panel). The size of the dots is a measure of the ionizing emissivity of the sources. Both panels show the same slice of the simulation volume. The thickness of the slice is 312.5~ckpc$/h$. The AGN are randomly hosted in 1\%  of the most massive haloes in the simulation volume. After 10~Myr, these AGN are turned off, and a different set of haloes are chosen to host new AGN. This can be seen as the change in the position of the red markers between the two panels. The galaxy distribution remains the same in both panels, as it is updated every 40~Myr.}
  \label{fig:quasar_dist}
\end{figure*}
The AGN in the simulations are assumed to be transient with a lifetime of 10~Myr. After this period, AGN stop emitting ionizing photons, and a new set of haloes is chosen to host AGN. At any given redshift, the combined emissivity of the AGN in the QSO-assisted model is chosen to be a fraction of 20\% of the stellar ionizing emissivity. The AGN are assumed to emit ionizing radiation at a redshift from $z=10$ in our QSO-assisted model. The haloes/galaxies chosen to host AGN are randomly selected from the most massive haloes in the simulation. Once the halos are selected, the AGN ionizing emissivity is proportional to the host halo mass. See  Appendix~\ref{sec:mass_distribution} for further details.

A representation of the spatial distribution of sources is shown in Figure~\ref{fig:quasar_dist} for our fiducial model for a slice of 0.78125~cMpc$/h$ thickness. The blue dots represent the galaxy-like sources, while the red dots represent the AGN-like sources. The sizes of the dots represent the ionizing emissivity of the sources. The left curve of the Figure is at $z=6.10$, while the right curve is at $z=6.05$. The time difference between the two panels is equal to 10~Myr. Note that the galaxy distribution in both panels is the same, as it is taken from the hydrodynamical simulation and is updated every 40~Myr. A new halo set is chosen to host AGN at the end of the 10~Myrs (at $z=6.05$). This is demonstrated by the red dots changing between the two panels. We further see that, on average, the most massive haloes/brightest galaxies are chosen to host AGN.

\subsection{The simulation suite}\label{sec:sim_suite}

The following sections will discuss the results from four different simulations. The first two are multi-frequency hydrogen and helium galaxy-only models without AGN from \citet{Asthana2024}. The spectrum of the ionizing photons produced by massive stars in the galaxies is modelled as a black-body with a temperature of 40000~K. The spectrum is divided into four different frequency bins. The two models have a relatively late start to reionization.  All models considered here  assume a linear scaling of the ionizing emissivity with halo mass. For the emission by massive stars in  the fiducial model 
and the QSO-assisted model this extends to the lowest mass haloes with  $10^9$~M$_\odot/h$.
In the Oligarchic model, the minimum halo mass hosting massive stars that produce ionizing photons is $8.5\times10^9$~M$_\odot/h$. The Oligarchic model is meant to explore the effects of rarer sources hosted only in more massive haloes (see also \citeyearname{Chardin2015}, and \citeyearname{Cain2023}).

The other two models host AGN. The `QSO-assisted' model assumes that 17\% of the total ionizing emissivity is due to accretion onto supermassive black holes in faint AGN and is described in detail in Section~\ref{sec:quasar}. The number of faint AGN at any given time in the QSO-assisted simulation is taken to be 1\% of the total number of haloes hosting galaxies producing ionizing photons. The lifetime of the AGN is 10~Myr. The AGN start emitting ionizing photons at $z=10$. The fourth model is a `QSO-only' model in which the fraction of haloes  hosting AGN is taken to be 10\%, and the AGN are emitting from $z=17$. There are no galaxy/stellar ionizing sources in the QSO-only model.

\subsection{Calibration to the \lya\ forest at $5\lesssim z\lesssim6.2$}\label{sec:calibrate}
We randomly extract 6,400 distinct sight-lines  from each simulation snapshot. The spatial resolution of the sight-lines and the density grid used by \atmf\ are the same. We calculate mean flux and the optical depth distribution. The approximation to the Voigt profile as described in \citet{Thorsten-2006} is used when computing \lya\  optical depths. The effective optical depth $\tau_\mathrm{eff}$ is calculated by the conventional definition, $\langle F\rangle=e^{-\tau_\mathrm{eff}}$, measured in $50\:\mathrm{cMpc}/h$ segments of the \lya\ forest. The mean flux computed through this method is calibrated to match the measurements from observed QSO absorption spectra presented in \citet{Bosman-2022}. This is done by modulating the ionizing volume emissivity in the simulation until the mean flux agrees with the observations for redshifts $5\lesssim z \lesssim 6.2$. Multiple iterations are required to fit the mean \lya\ forest transmission well. A separate calibration was done for each model, using the  approach described above.

\begin{table*}
  \centering
  \begin{tabular}{|c|c|c|c|c|c|c}
    \hline
    Simulation & Minimum halo mass & AGN & AGN share of & Redshift of & Fraction of & AGN \\     
    & for sources & contribution & H-ionizing emissivity & AGN onset & haloes with AGN & lifetime \\
    \hline
    Fiducial &  $10^9\,$M$_\odot/h$& No& --&--&-- &--\\
    Oligarchic&  $8.5\times10^9\,$M$_\odot/h$& No&--&--&--&--\\
    QSO-assisted &$10^9\,$M$_\odot/h$&Yes& 17\% & 10 & 1\%&10~Myr\\
    QSO-only  &$10^9\,$M$_\odot/h$&Yes& 100\%& 17 & 10\%&10~Myr\\
    \hline
  \end{tabular}
  \caption{Simulations presented in this paper. In each case, the simulation box size is 160~cMpc$/h$, with a $2048^3$ uniform Cartesian grid for the radiative transfer, corresponding to a spatial resolution of $78.125$~ckpc$/h$. The relative contribution of AGN is independent of redshift.\label{table:simulation_suite}}
\end{table*}

\section{The effect of faint AGN on reionization} \label{sec:QSOassist}
\begin{figure*}
  \centering
  \includegraphics[width=\textwidth]{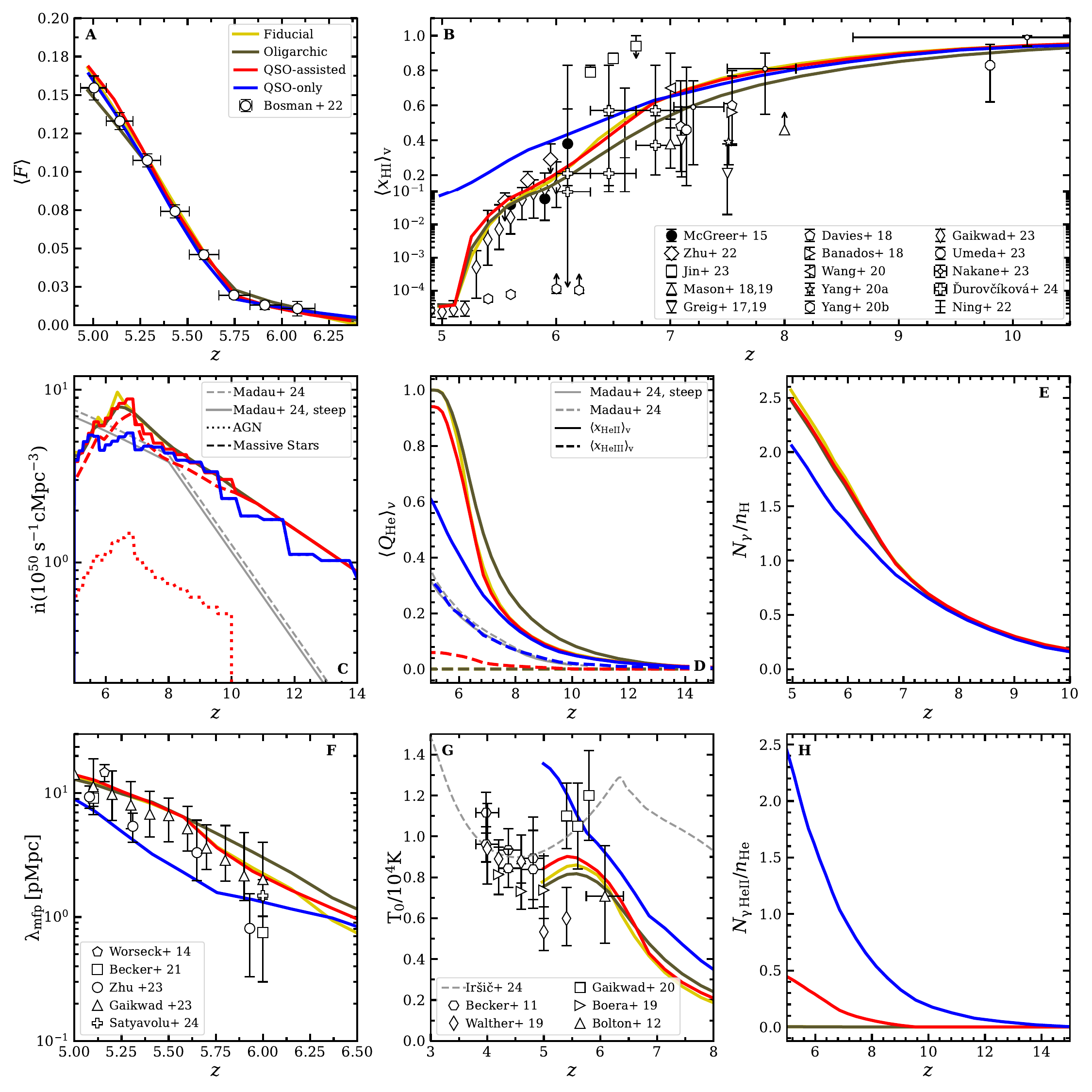}
  \caption{Various quantities related to reionization from the four simulations discussed in the paper. Panel~A compares the  mean \lya\ forest transmission, $\langle F\rangle$, with measurements by \citet{Bosman-2022}.  Panel~B shows the volume-averaged neutral hydrogen fraction, $\langle x_\mathrm{HI}\rangle_\mathrm{v}$.  This panel also shows inferences of the neutral hydrogen fraction from various observations, the fraction of Lyman-break galaxies showing \lya\ emission \citep{Mason-2018, Mason-2019}, dark gaps in the \lya\ forest \citep{Mcgreer-2014, Zhu-2022, Jin-2023}, \lya\ emission equivalent widths \citep{Nakane-2023}, quasar damping wings \citep{Greig-2017, Banados-2018, Davies-2018, Greig-2019, Wang-2020, Yang-2020, Durovcikova-2024}, the effective \lya\ opacity of the IGM \citep{Yang-2020-2, Ning2022, Gaikwad-2023}, and galaxy damping wings \citep{Umeda-2023}. Panel~C shows the total ionizing emissivity. For the  QSO-assisted model, the emissivity contribution by faint AGN  and massive stars in galaxies are shown separately by the dotted and dashed lines, respectively. The solid curve shows the total emissivity. The solid and dashed grey curves show the emissivity of two QSO-only models from \citet{Madau2024}. Panel~D shows the volume-averaged ionized fractions $\langle x_{\text{HeII}}\rangle_\text{v}$ and  $\langle x_{\text{HeIII}}\rangle_\text{v}$, as solid and dashed curves, respectively. Panel~E shows the ratio of the integrated number of hydrogen-ionizing photons to the number of hydrogen atoms. Panel~F compares the mean free path of hydrogen-ionizing photons with measurements by \citet{Worseck-2014, Becker2021, Zhu-2023, Gaikwad-2023}, and \citet{Satyavolu-2023-b}.  Panel~G shows the evolution of the mean IGM temperature at mean density to measurements by \citet{Becker2011, Bolton2012, Boera-2019, Walther-2019, Gaikwad-2020}, and the best guess model from \citet{Irsic2024}. The dashed coloured curves shows the mean temperature of the ionized gas at mean density. Panel~H shows the ratio of the integrated number of \heii\ ionizing photons to the number of helium atoms.}
  \label{fig:mosaic_quasar}
\end{figure*}

Our QSO-assisted and QSO-only models allow us to investigate the impact of faint AGN on reionization. In Figure~\ref{fig:mosaic_quasar}, we compare a range of critical diagnostics for the two new models with two galaxy-only models from \citet{Asthana2024}, the fiducial and Oligarchic models. The fiducial, Oligarchic, QSO-assisted and QSO-only models are shown by yellow, brown, red and blue curves, respectively. 

\subsection[]{Ionizing volume emissivity}\label{sec:ion_emi}
In panel~A of Figure \ref{fig:mosaic_quasar}, we plot the evolution of the mean \lya\ forest transmission as a function of redshift. The ionizing emissivity in all four models has been calibrated to fit the measurements by \citet{Bosman-2022} as described in Section~\ref{sec:calibrate}. The corresponding ionizing volume emissivity is shown in panel~C of Figure \ref{fig:mosaic_quasar}. All models have a rising emissivity up to about $z\sim 7$, after which the emissivity declines. The steepness of the drop required to match the \lya\ transmission is somewhat model-dependent. 

The evolution of the ionizing emissivity in the Oligarchic model is similar to that in the fiducial model, but with a slightly lower peak. The emissivity in the QSO-assisted model is the same as in the fiducial model up to $z=10$, after which the faint AGN are assumed to start ionizing photons. While the faint AGN contribute, the total ionizing emissivity required to match the observed \lya\ transmission is similar to that in the galaxy-only models.. The solid pink curve is the total emissivity by faint AGN  and massive stars in galaxies. In contrast, the dashed and dotted curves are the contribution of the massive stars in the galaxies and the faint AGN hosted by the galaxies in the QSO-assisted model. We have set the AGN  contribution to be 20\% of the ionizing emissivity of the massive stars in the galaxies across all redshifts. The choice of 20\%  is motivated by the discussion of the AGN fraction in \citet{Maiolino2023b} and corresponds to 17\% of the total ionizing emissivity.  As discussed in \citet{Cain2024}, the physical reason for the decline in emissivity is not clear. Some discussion of this drop can also be found in \citet{Qin-2021}, which do not require such a drop in their simulations with the semi-numeric code  21~\textsc{cmfast}. \citet{Ocvirk-2021} have argued that radiative suppression of star formation could (partially) explain such a drop. A decreasing escape fraction due to increasing metal and dust content produces a more gentle decrease in the \textsc{sphinx} simulation \citep{Rosdahl2022}. An earlier start with a more gradual reionization history also creates a smaller and more gentle drop in the Early and Very Early models in \citet{Asthana2024} (see also  \citeyearname{Cain2024-2}).

In the QSO-only model represented by the blue line, the ionizing emissivity of massive stars in galaxies is set to zero. This model requires the lowest ionizing emissivity of the four models discussed here to match the mean observed \lya\ transmission. We also have overplotted the emissivity of two QSO-only models advocated by \citet{Madau2024}. The two models assume two different combinations of the spectral index for the ionizing emissivity and \heii\ escape fractions of the QSOs, chosen so that not too much \heii\ get ionized to \heiii\  by $z\sim6$ (see appendix~\ref{appen:madau_cal} for further details). The ionizing emissivity of both these QSO-only models is similar, but somewhat higher than in our QSO-only model below $z\sim 8.5$, and does not show a drop below $z<6$. With our simulations such an increasing emissivity at $z<6$ will not fit the \lya\ transmissivity at $5<z<6.2$.

Panel~E shows the integrated number of photons per hydrogen atom $N_\gamma/n_H$ required for reionization as a function of redshift. The integrated number of photons is computed directly from the volume emissivity. In the QSO-only model the ionizing volume emissivity and the $N_\gamma/n_H$ are somewhat lower. This is probably due to a combination of reionization occurring somewhat later,  a smaller mean photo-ionization rate required to match the mean \lya\ forest transmission, a slightly higher temperature reducing recombination rates, the ionizing luminosity being enhanced for 10~Myr during the AGN phase allowing ionizing photons to escape with fewer recombinations in the host halo, and the increased patchiness due to the possibility that low mass haloes can also host an AGN. In panel~H, we show the integrated number of \heii\ ionizing photons per helium atom as a function of redshift. The galaxy-only models are close to zero due to the lack of \heii\ ionizing photons. Note that the about a factor six lower number of \heii\ ionizing photons in the QSO-assisted model compared to the QSO-only model results in a \heiii\ volume factor that is  a factor six lower as expected.     

\begin{figure*}
  \centering
  \includegraphics[width=1.6\columnwidth,page=1]{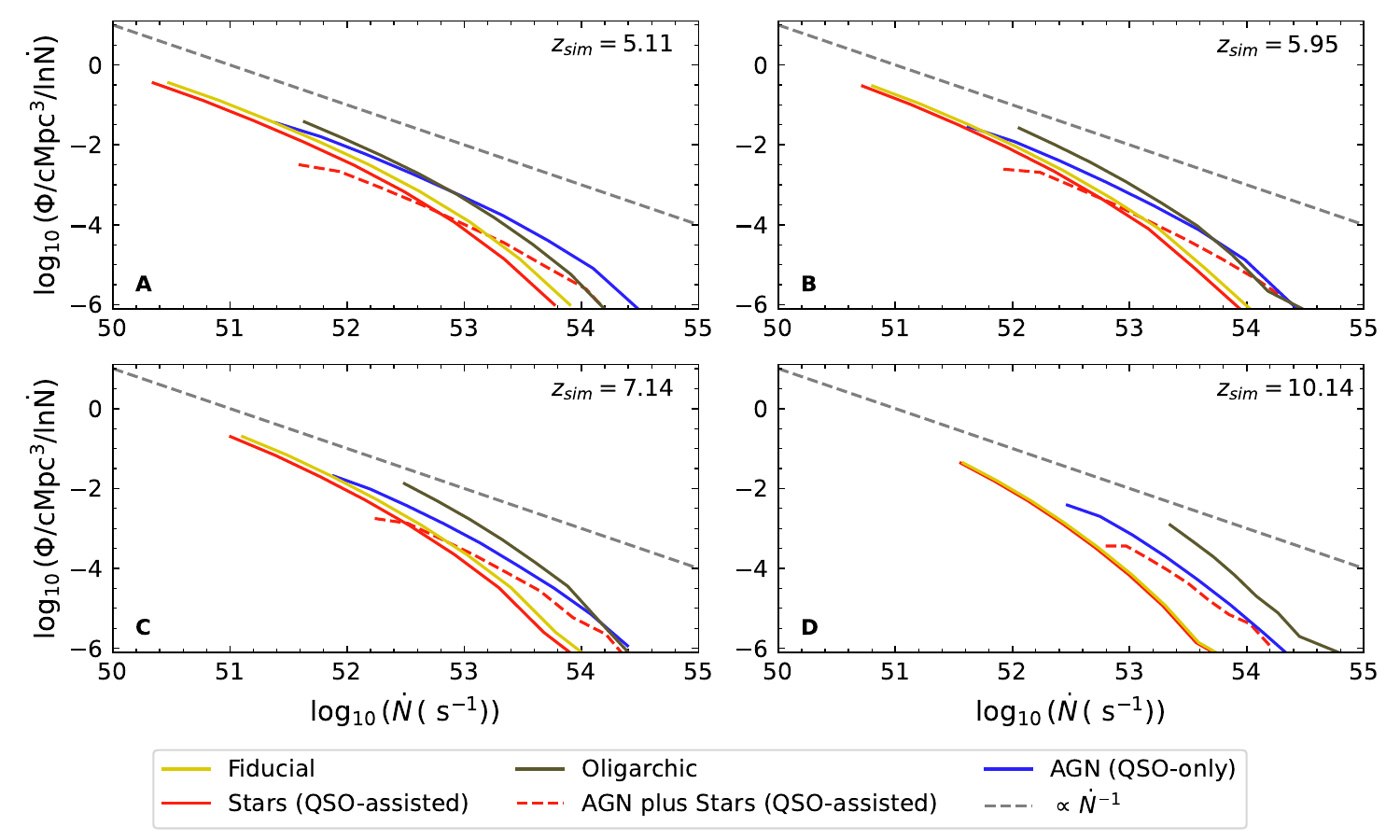}
  \caption{The hydrogen-ionizing luminosity function of the  sources in our simulations. The solid red curve represents the ionizing emissivity in the QSO-assisted model due to stars in galaxies that do not host an AGN. The dashed red curve shows the distribution for AGN plus stars in galaxies that host an AGN. The blue curve represents the distribution for the QSO-only model, while the yellow and brown curves represent the fiducial and Oligarchic galaxy-only models. }
  \label{fig:hist_ndot}
\end{figure*}

\subsection{Reionization history} \label{sec:reionization_history}
In panel~B of Figure~\ref{fig:mosaic_quasar}, we show the evolution of the volume-averaged neutral fraction as a function of redshift overplotted with observational estimates from measurements of the fraction of Lyman-break galaxies showing \lya\ emission \citep{Mason-2018, Mason-2019}, dark gaps in the \lya\ forest \citep{Mcgreer-2014, Zhu-2022, Jin-2023}, \lya\ emission equivalent widths \citep{Nakane-2023}, quasar damping wings \citep{Greig-2017, Banados-2018, Davies-2018, Greig-2019, Wang-2020, Yang-2020, Durovcikova-2024}, the effective \lya\ opacity of the IGM \citep{Yang-2020-2, Ning2022, Gaikwad-2023}, and galaxy damping wings \citep{Umeda-2023}. Our QSO-assisted and the two galaxy-only models from \citet{Asthana2024} fall well within the observational estimates. Our new QSO-only model ionizes significantly later than the neutral fraction evolution derived from \lya\ forest data despite the model fitting the evolution of the mean \lya\ forest transmission. We will return to this apparent contradiction in Section~\ref{sec:AGN_ion}, where we compare the models in more detail. 

The Oligarchic model has a similar ionizing emissivity at high redshift as the fiducial galaxy-only model but has a lower ionized fraction. This suggests that sources hosted in more massive haloes are more efficient in keeping hydrogen ionized. In order to slow completing reionization, the emissivity requirement for the Oligarchic model decreases compared to the fiducial towards the end of reionization. This is also seen in the QSO models. The emissivity required to complete reionization is a factor 1.8 lower in the QSO-only models as discussed in Section \ref{sec:ion_emi}. Note that at high redshift the ionizing emissivity is still extremely uncertain. Thus, for the QSO models we have chosen emissivity values similar to our galaxy-only models for ease of comparison.


In panel~D of the figure, we show the evolution of the volume-averaged fractions of singly ionized helium (solid) and doubly ionized helium  (dotted). In the two galaxy-only models, the \heii\ fraction closely follows the hydrogen fraction as expected without hard photons, as \heiii\ is not ionized. With AGN emitting hard photons capable of ionizing \heii, by $z=5$ volume-averaged  6\% \heii\ has been ionized to \heiii\ in the QSO-assisted model and  in the QSO-only model this fraction is 35\%. Note that the fraction in our new QSO-only model is similar to that in the two QSO-only models in \citet{Madau2024} and is consistent with \heii\ reionization completing by $z\sim 3$ \citep{Furlanetto2010, Worseck2016, Puchwein2019, Gaikwad2021, Makan2021}. 

In panel~G of Figure~\ref{fig:mosaic_quasar}, we compare the evolution of the mean temperature at mean density with redshift for our models\footnote{Note that we assume photo-heating rates in the optical thin limit (see \citeyearname{Asthana2024}) which may somewhat underestimate the temperatures in just ionized gas.}  with a collection of temperature measurements by \citet{Bolton2012, Garzilli-2017, Walther-2019, Boera-2019, Gaikwad-2020, Irsic2024}. Our models are in broad agreement with the data. However, note that the observed values are not consistent with each other and more data is needed as well 
as a wider range of models. The peaks in the curves correlate closely with the reionization history. The QSO-assisted model has a slightly higher peak in temperature due to the presence of harder photons. The QSO-only model has a rising temperature curve because reionization is incomplete. It also has the highest temperature as the number of harder photons in the QSO-only model are larger than the QSO-assisted model leading to a larger fraction of \heii\ being ionized to \heiii.

Panel~F of the same figure shows the average mean free path of ionizing photons as described in \citet{Kulkarni2016}. The mean free path  is calculated by averaging the Lyman-continuum transmission from random sight-lines in the comoving frame and fitting an exponential with an e-folding length scale of $\lambda_\mathrm{mfp}$,
 \begin{equation}
   \langle \mathrm{exp}(-\tau_{912}) \rangle = F_0\, \mathrm{exp} \left( -\frac{x}{\lambda_{\mathrm{mfp}}} \right), 
 \end{equation}
 where $\lambda_\mathrm{mfp}$ is the mean free path, and $x$ is the position along a sight-line \citep{Rybicki-1986}. Except for the QSO-only model, the simulations have similar mean-free paths. The  differences can be traced back to differences in neutral fractions. The calculated mean free path from the models agrees well with that measured by \citet{Gaikwad-2023}. However, they are a factor two larger than those measured in the near-zones of QSOs at $5.8<z<6$ by \citet{Becker2021}, \citet{Satyavolu-2023-b}, and \citet{Zhu-2023}. At lower redshift, the models and observations converge. Furthermore, as expected, the volume-averaged neutral fraction and the mean free path are anti-correlated.  As a consequence, the mean free path of the QSO-only model, where reionization completes later, has a lower mean free path at $5<z<6$  and is within the error bars of the measurements by \citet{Becker2021} at $z\sim 6$, but falls below the measurements at lower redshift. 
 
\subsection{Luminosity function of ionizing sources}
As discussed earlier, the volume ionizing emissivity was chosen to match the observed mean \lya\ optical depths and then distributed over the dark matter haloes identified in the simulation as described earlier. 
The corresponding luminosity function of ionizing emissivities (in ionizing photons emitted per second) is shown in Figure~\ref{fig:hist_ndot}. The four models are indicated by the same colours as the previous sections, i.e. yellow, brown, red and blue for the fiducial, Oligarchic, QSO-assisted, and QSO-only models. The dashed red curve represents the AGN and their host galaxies, while the solid red curve represents those without AGN. 

In panel~D, the Oligarchic model has significantly brighter and fewer fainter ionizing sources than the fiducial model. This is expected as in the Oligarchic model, the halo mass cutoff to host massive stars emitting ionizing photons is higher. This trend is seen across all redshifts. In the QSO-assisted model, the distribution is very similar to the fiducial model but shifted to the left. This is expected as the required emissivity requirement is lower.
Furthermore, the brightest AGN are tenfold brighter than the brightest ionizing sources powered by the massive stars in galaxies at all redshifts. 
The QSO-only model has a higher number of brighter and fainter AGN than the QSO-assisted model. This is also expected as QSOs comprise the entire ionizing emissivity in this model, and the halo occupation fraction of massive haloes is ten times larger on average.

\subsection{Effective \lya\  optical depth distribution}
\begin{figure}
  \centering
  \includegraphics[page=1, width=\columnwidth, trim={0.3cm 0.2cm 0.5cm 0.2cm},clip]{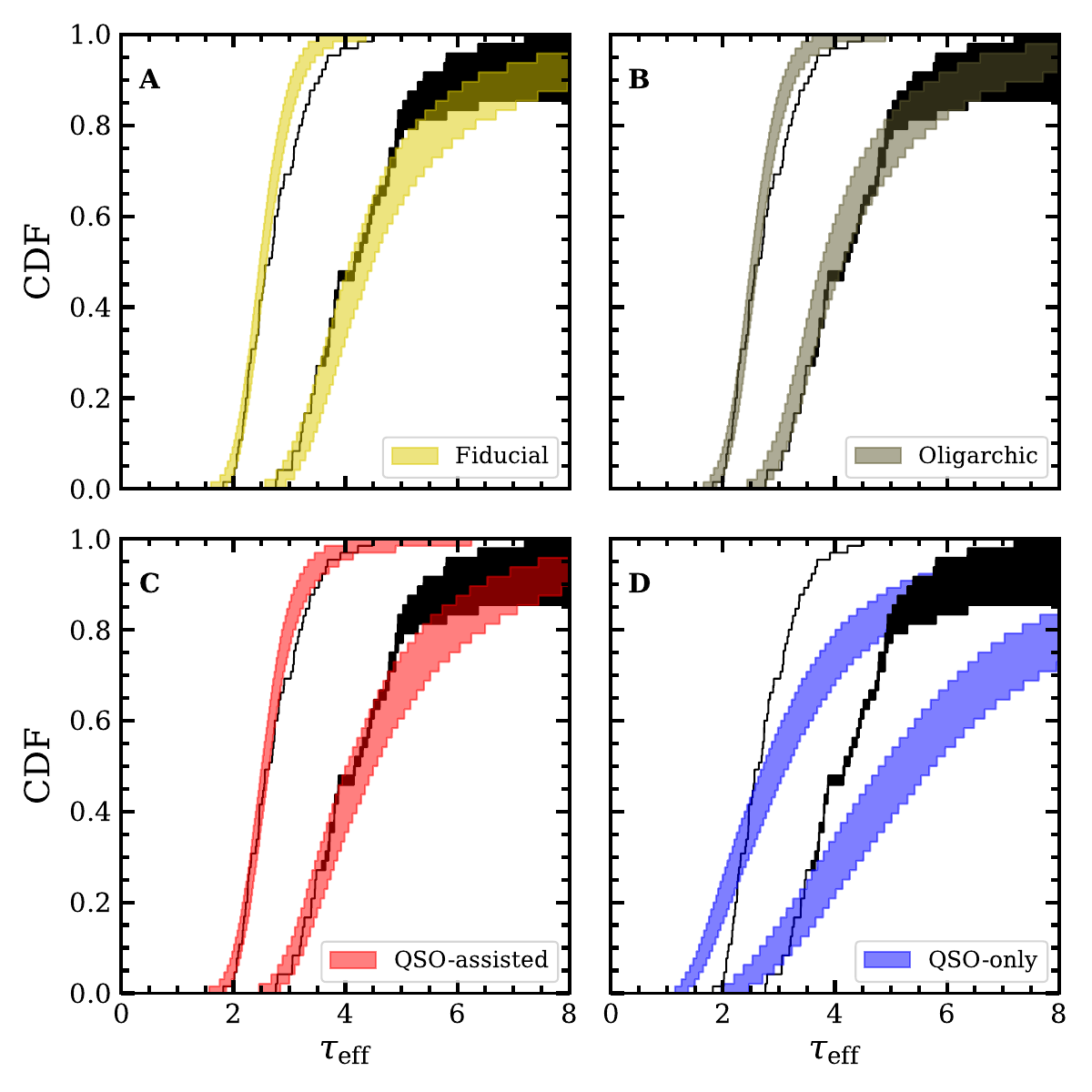}
  \caption{Panels~A--D show the cumulative distribution functions (CDFs) of the effective \lya\ opacity $\tau_\mathrm{eff}$ of the IGM computed over 50~cMpc$/h$ sections of the \lya\ forest for the four models presented in this paper.  Each of these panels shows two CDFs. The CDF with smaller average values of $\tau_\mathrm{eff}$ corresponds to  $5.36<z<5.51$, and the other CDF corresponds to $5.67<z<5.83$. The shaded regions denote the 16\%-84\% percentiles of 10,000 realizations of the CDF, each created using the same number of lines of sight as in the observed sample, i.e., 65 and 48, respectively, at the two redshifts. The black shaded regions in each of the panels show measurements by \citet{Bosman-2022}.}
  \label{fig:mosaic_cdf_quasar}
  \end{figure}
As already discussed in detail in \citet{Asthana2024} for galaxy-only models, calibrating to the mean \lya\ flux still leaves room for variations in the width of the cumulative distribution function of the effective optical depth. In Figure~\ref{fig:mosaic_cdf_quasar}, we show  the cumulative distribution function (CDF) of the effective optical depth for the four models discussed in the paper. The models are shown in the same colours as in Figure~\ref{fig:mosaic_quasar}. The CDFs are calculated over $50\: \mathrm{cMpc}/h$ segments and are shown for  two redshift ranges, $5.36$ to $5.51$, and $5.67$ to $5.83$. The shaded regions denote the $15.87 - 84.13$ percentile of $10,000$ realizations of the CDF, each created using the same number of lines-of-sight (LOS) as the observed data, i.e., $65$ and $48$, respectively. The black curves show the observational data from \citet{Bosman-2022}. 

Focusing on panel A-C, we see that calibrating to the mean flux does not necessarily translate into reproducing the width of the observed distribution, especially at the tail of high optical depths. As discussed in \citet{Asthana2024}, this is because once the effective optical depth in a $50$~cMpc$/h$ segment of the spectrum is significant, the exact value has little effect on the median of the effective optical depth distribution.  

Looking in more detail at the redshift range  $z=5.36$ to $5.51$ (left set of curves in each panel), we see that the QSO-assisted model fits the data well with a somewhat more extended tail towards high optical depth. This is because in this model, reionization ends somewhat later, and neutral islands persist slightly longer than in the galaxy-only models; thus, it has enough neutral cells at $z\sim5.4$ to get the correct width of the CDF at high values of the optical depth while the other models do not. Progressing to the redshift bin of $5.67<z<5.83$, the large effective optical depths in this redshift range represent the last remaining neutral islands, and the galaxy-only models appear to have fewer of these in the redshift range $5.67<z<5.83$. In the QSO-only model, the width of the distribution  at both redshifts is larger than the observations. 

By introducing our models with faint AGN into the comparison, the differences in the optical depth distribution and the evolution of the neutral hydrogen fraction for models that fit the mean \lya\ forest transmission become much more significant. The inferred neutral hydrogen fraction is thus model dependent, and the error bars in the measurements of \citet{Gaikwad-2023} likely underestimate the total uncertainty. 

\section{Results and Discussion} \label{sec:discussion}

\subsection{The effect of faint AGN on ionized \hii\ regions}\label{sec:AGN_ion}
\begin{figure*}
  \centering
  \begin{minipage}{1.0\textwidth}
    \includegraphics[width=1\textwidth,page=1,trim={5.5cm 1.4cm 5cm 0cm},clip]{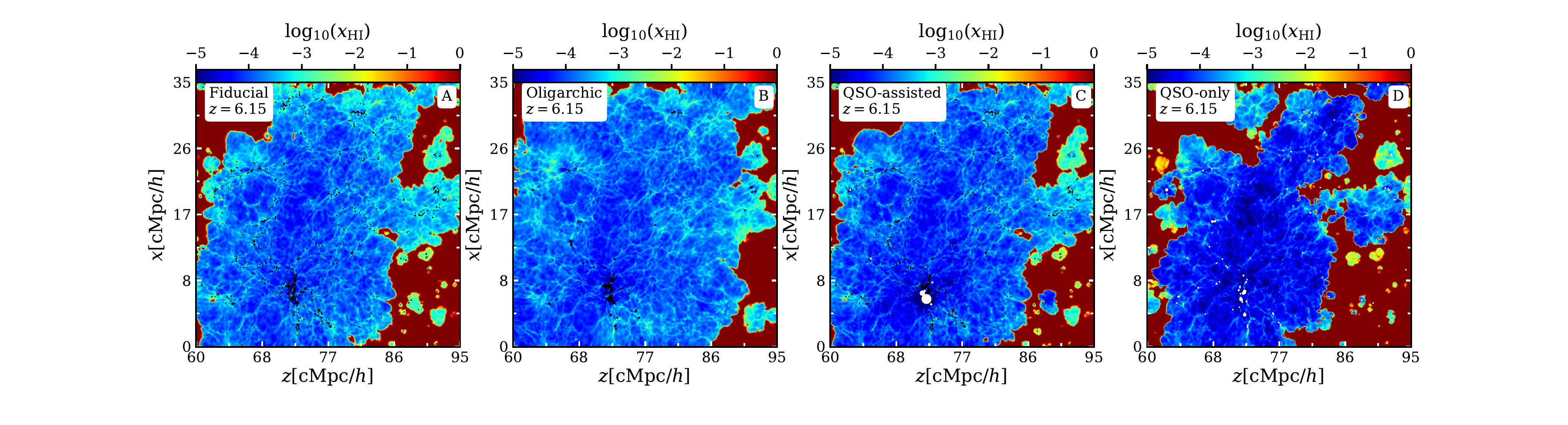}  
  \end{minipage}
  \begin{minipage}{1.0\textwidth}
  \vspace{-0.2em}
  \includegraphics[width=1\textwidth,page=1,trim={5.5cm 1.4cm 5cm 0cm},clip]{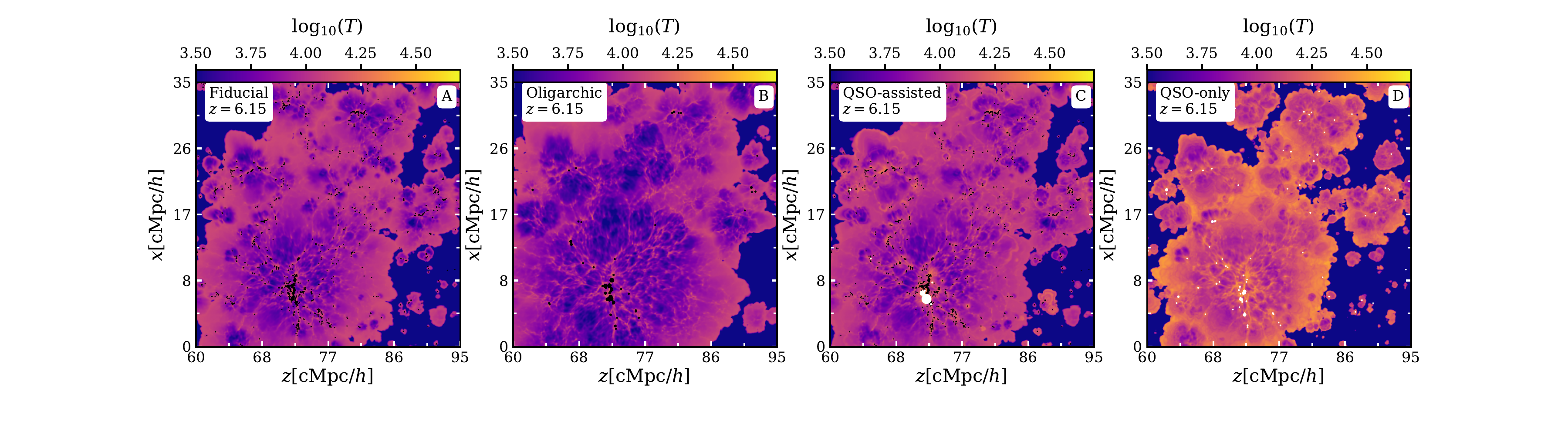}
  \end{minipage}
  \caption{Spatial distribution of the ionized hydrogen fraction (top row) and gas temperature (bottom row) in the four simulations presented in this paper at $z=6.15$.  Each slice has a thickness of 78.125~ckpc$/h$; black markers represent galaxies, while white markers represent AGN. The size of the markers is proportional to the ionizing luminosity of the sources. }\label{fig:slice_plots}
 \end{figure*}

In the previous section, we showed that the QSO-assisted model agrees well with the observed effective
optical depth distribution. We had already discussed that the evolution of the neutral fraction in the QSO-assisted model is different from that in the galaxy-only models and that reionization completes somewhat later with neutral islands remaining present for longer,  improving the agreement at high effective optical depth. However, there are also differences in the neutral fraction 
in ionized regions, resulting in low effective optical depth as shown in Figure~\ref{fig:slice_plots}. Here, we show the ionized hydrogen fraction and temperature of the IGM in a slice of the fraction of the simulation volume at $z=6.15$
for our four models. The black dots and their size represent the location and ionizing luminosity of the galaxies, while the white dots represent that of the  QSOs. The slice is $39~\text{cMpc}/h$ across, and the thickness of the slice showing the IGM properties is 78~ckpc$/h$. The ionizing sources are plotted in projection for a thicker slice with a width of 150~ckpc$/h$ to catch a larger number of sources in the slice.

At the bottom of the slice, a cluster of sources is visible. In the QSO-assisted model, this region has a lower value of the neutral hydrogen fraction than the galaxy-only model. This same region is hotter, as seen in the temperature maps. In the QSO-only model, this effect of the faint AGN is more pronounced, as seen by the overall lower neutral fraction in the ionized regions. However, in the QSO-only model, the overall ionized region is smaller, and the small ionized bubbles are not as ionized as in the fiducial model. This is an effect of the smaller number of sources in this model, and that reionization is completed later. 

The larger number of \heii\ ionizing photons are leading to a significantly higher temperature in the region immediately 
surrounding an AGN and in areas recently exposed to the hard photons by previous generations of our transient AGN that emit for 10~Myr. This additional heating with hard photons of AGN is also reflected in the higher average temperature at mean density that we discussed in Section~\ref{sec:reionization_history}. 

Along any line-of-sight, most of the contribution to the mean flux  comes from the ionized cells. As the region around the QSOs is highly ionized, the contribution to the mean flux of these cells is larger. 
Along the rest of the line-of-sight, hydrogen can thus nevertheless be more neutral in simulations calibrated to the same mean flux as in the galaxy-only models. 


In the QSO-only model,  due to the smaller number of AGN, the ionizing emissivity of each source is much larger. This leads to more highly ionized regions around the AGN and a larger neutral fraction away from the sources for  models that  match the observed mean \lya\ forest transmission. Due to the similar distribution of the ionized regions, the QSO-only model reproduces the same mean \lya\ forest transmission, but the effective optical depth distribution is too broad. In Appendix~\ref{appen:pdf} we compare  the pdf of the neutral fraction
for the different models. 

\subsection{Differences between Oligarchic galaxy-only and the QSO-assisted model}
In the Oligarchic model, because of the higher minimum  mass ($8.5\times10^9$~M$_\odot$) of haloes hosting ionizing sources, high-mass haloes host sources  with larger ionizing luminosity. The QSO-assisted model is similar, but with a much more significant difference between the emissivity of sources hosting AGN and the ones without. Furthermore, the spectra are different with a much larger number of hard photons in the QSO-assisted model. Clear differences between the two models become apparent  on closer inspection of the various panels of Figure~\ref{fig:mosaic_quasar}. In panel~B, reionization in the QSO-assisted model ends later, the IGM is slightly hotter, and there is potentially better agreement with regard to the width of the optical depth distribution compared to the Oligarchic model the lower redshift bin. Significant differences are visible in the neutral hydrogen fraction in the ionized bubbles, as shown in Figure~\ref{fig:slice_plots} and Appendix~\ref{appen:pdf}. Comparing panel~A and panel~B, we can see that due to the  brighter sources in the Oligarchic model the IGM is more highly ionized close to the bright sources and less highly ionized further away. As the minimum halo mass hosting ionizing sources is larger, the ionized regions produced by the sources in low-mass haloes in the fiducial model are missing in the Oligarchic model. Incorporating AGN creates both large and small ionized regions. Including AGN furthermore leads to a more significant contrast between the ionized and neutral regions. 

\subsection{Early reionization of \heiii\ by faint AGN}
\begin{figure*}
  \centering
  \includegraphics[page=1,width=1.0 \textwidth,trim={5.5cm 2.8cm 5cm 0cm},]{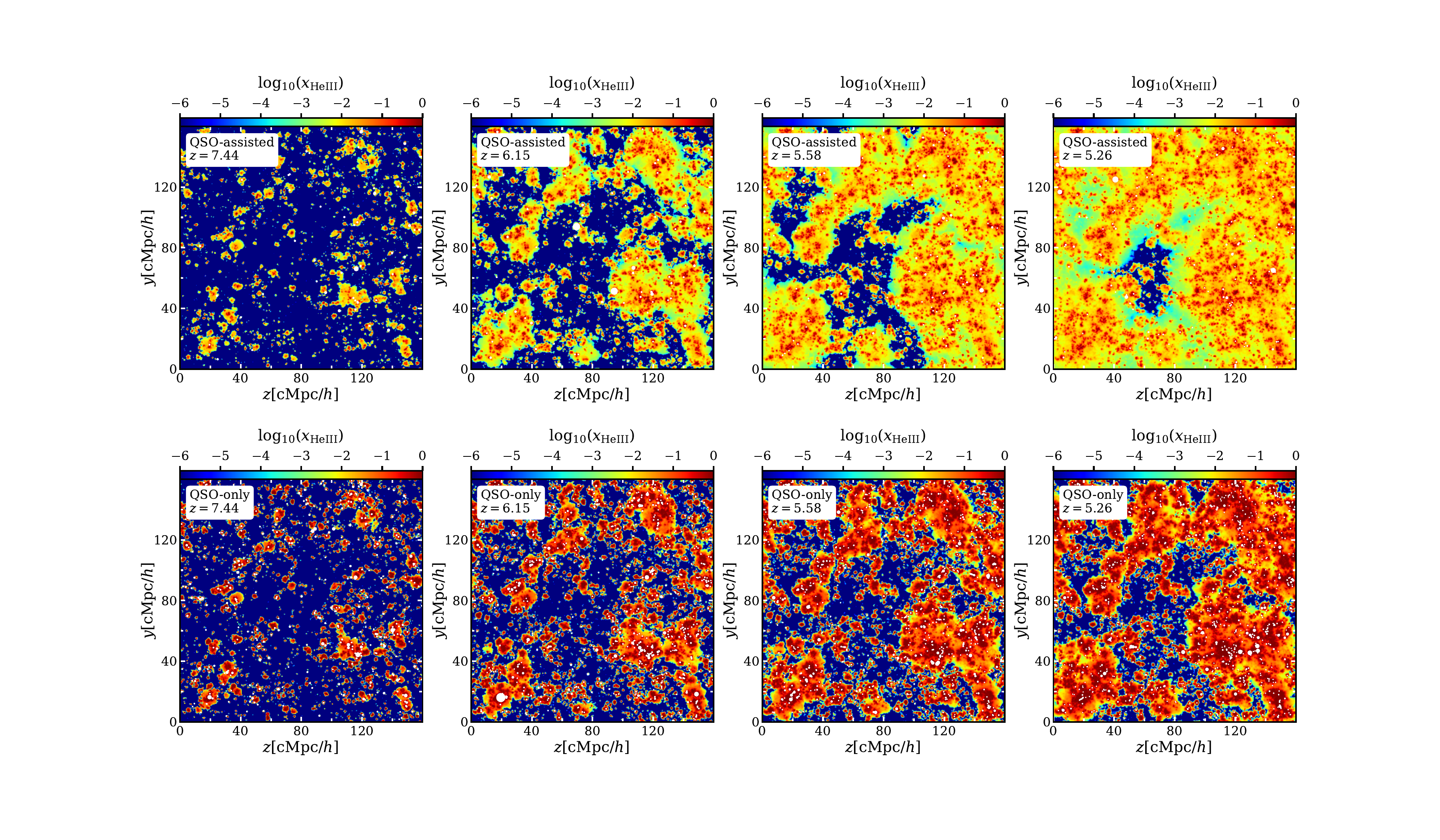}
  \caption{Spatial distribution of the \heiii\ fraction in the QSO-assisted model (top row) and the QSO-only model (bottom row)at $z\,=\,7.44$, 6.15, 5.58 and 5.26. Each slice has a thickness of $78.125$~ckpc/$h$ and a width of $160$~cMpc/$h$. The white dots represent AGN in the top and bottom rows. The AGN are randomly distributed in the simulation volume. The sizes of the dots are a measure of the ionizing luminosity of the sources. }
  \label{fig:slice_plots_HeIII}
\end{figure*}
The harder photons emitted by the faint AGN  lead to significant ionization of \heii\ to \heiii. In Figure~\ref{fig:slice_plots_HeIII}, we show slice plots of the \heiii\ fraction at four different redshifts ($z=7.44$, 6.15, 5.58, 5.26) for the QSO-assisted model in the top row, and the QSO-only model in the bottom row. The white dots and their size represent the location and ionizing emissivity of the faint AGN. The slice is $160~\text{cMpc}/h$ across, and the thickness of the slice showing the IGM properties is 78~ckpc$/h$. The ionizing sources are plotted in projection for a thicker slice with a width of 150~ckpc$/h$ to catch a larger number of sources.

In the QSO-assisted model (top row of the figure) the ionization of \heii\ is progressing similar to \hi. Ionized bubbles form around the faint AGN, which merge as the simulation proceeds. At the location of the AGN, the \heiii\ fraction temporarily reaches unity in the dark red regions surrounding the white dots. However, due to the relatively short lifetime of 10~Myr of the faint AGN and the high recombination rate of \heiii\ (about 5 times higher than that of  \hii), these regions do not stay fully ionized. This causes most of the volume of the \hii\ regions, which coincide with the \heii\ regions, to be partially ionized to \heiii\ while the areas close to  the AGN (currently emitting ionizing photons) are temporarily completely ionized. 

The contribution to the emissivity by faint AGN is 20\% of the contribution of the massive stars in the galaxies in the QSO-assisted model. In the QSO-only model, all the ionizing photos  are emitted by faint AGN. This leads to a factor of about six larger numbers of \heii\ ionizing photons. In panel D of Figure~\ref{fig:mosaic_quasar}, the \heiii\ fraction is 30\% in the QSO-only, a factor of six larger than in the QSO-assisted model. Naively, one may expect a similar picture in the QSO-only as in the QSO-assisted model, i.e. large ionized bubbles of partially ionized \heii, but with a few regions with complete ionization of \heii. However, this is not what is seen in the bottom row of the figure. First, note that the number of white dots is larger than in the QSO-assisted model, which is expected as the number of galaxies hosting an AGN in this model is 10\% compared to the 1\% in the QSO-assisted model. We also see that the emissivity of these objects is larger (see Section~\ref{sec:UV_lum}). Around the AGN, as expected, we see a complete ionization of \heii\ (the red regions). Due to the lifetime and high recombination rate, these regions do not stay ionized, similar to the QSO-assisted model. However, the overall size of the \heiii\ regions is much smaller. This is because there are no pre-existing \hii\ and \heii\ regions when AGN turns on due to the absence of ionising photons emitted by massive stars in galaxies. The dark blue regions in the bottom row are regions where \hei\ ionization has not yet occurred. The more significant number of \heii\ ionizing photons in the QSO-only model creates thus a large number of completely ionized small \heii\ regions, but coalescence into large ionized \heiii\ bubbles proceeds slower than in the QSO-assisted model. Some of the hard photons in the QSO-only model are used up by  ionizing \hi\ and \hei.

\subsection{UV luminosity function and ionizing emissivity of galaxies and AGN}\label{sec:UV_lum}

\begin{figure*}
  \centering
  \includegraphics[width=\textwidth,page=1]{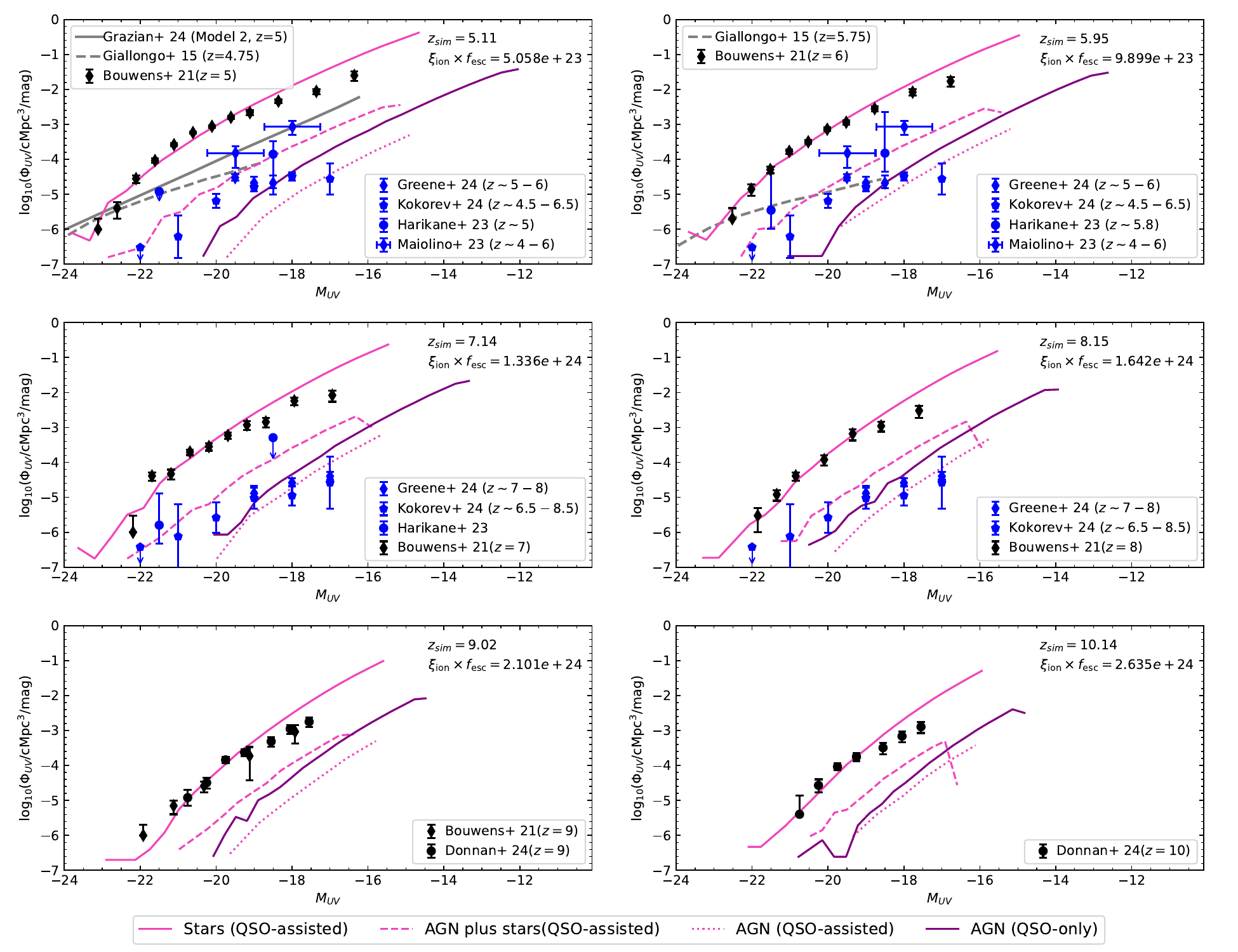}
  \caption{The inferred UV luminosity function of the stars in all galaxies, whether or not they host an AGN for the QSO-assisted model, is shown as the solid pink curve. A subset of these galaxies host AGN, and their UVLF with the UV luminosity from stars and the AGN combined is shown by the dashed pink curve. The dotted pink  shows the UVLF for the AGN contribution present in these galaxies. The redshifts at which the UVLF is computed are shown in the top right, followed by the values of   $ f_\mathrm{esc}\xi_\mathrm{ion}$ inferred from matching the  UVLF (solid pink curve) by eye to the observed data. Note that we assumed $f_{\rm esc}= 1$ for the fraction of hydrogen ionizing photons escaping from the faint AGN. This value is likely to be lower. For fixed ionizing luminosity the   UV$_{1450}$ luminosity of the AGN (contribution) scales as $f_{\rm esc}^{-1}$ and the luminosities thus could perhaps be up to about a magnitude brighter.  The solid purple line is the UVLF of the AGN from the QSO-only model. The observed UVLF for galaxies are taken from \citet{Oesch2018, Bouwens-2021, Donnan2024} and are plotted as black data points. The observed UVLF for AGN are taken from \citet{Giallongo2015, Maiolino2023b, Matthee2024, Harikane2023, Greene2024, Grazian2024, Kokorev2024}, and are plotted as blue data points.}
  \label{fig:uv_luminosity}
\end{figure*}

With further assumptions for the  SED of  the AGN population, we can also compute the UV luminosity function of these objects in our models.  We can furthermore relate UV to ionizing emissivity  for the galaxies in our models by assuming an ionizing photon production efficiency $\xi_\text{ion}$ and an escape fraction of ionizing photos $f_\text{esc}$. This allows us to compare the UV luminosity function of galaxies and AGN in our models with observations. We convert the luminosity to absolute magnitude by using the relation \citep{1983ApJ...266..713O},
\begin{equation}
  M_{\text{UV}} = -2.5\log_{10}\left(\frac{L_{\text{UV}}}{\text{erg s}^{-1}\text{Hz}^{-1}}\right)+51.63,
  \label{eq:UV_luminosity_2}
\end{equation}
where $M_\text{UV}$ is the absolute magnitude at 1450\,\AA, and $L_\text{UV}$ is the UV luminosity at the same wavelength. To calculate the UV luminosity of galaxies, we assume a value of escape fraction, $f_\text{esc}$, and ionizing photon production efficiency, $\xi_\text{ion}$, in units $\text{erg}\,\text{Hz}^{-1}$. This gives  $L_{\text{UV}} =\dot{N}/(f_\text{esc}\xi_\text{ion}$), where $\dot{N}$ is the hydrogen-ionizing emissivity from our simulation. Assuming  canonical values of $\log_{10}{\xi_\mathrm{ion}}=25.5$ and $f_\text{esc}=0.1$, we can write,
\begin{multline}
  M_{\text{UV}} = -19.62 -2.5\log_{10}\left(\frac{\dot{N}}{10^{53}\text{s}^{-1}}\right)\\+2.5\log_{10}\left(\frac{f_\text{esc}}{0.1}\right)+2.5\left(\log_{10}\xi_\text{ion}-25.5\right).
\end{multline}
For AGN, we assume a broken power law for $L_\text{UV}$,
\begin{equation}
  L_\text{UV} = f_\text{EUV}\left(\frac{\nu}{\nu_{912}}\right)^{\alpha_{\text{EUV}}},
  \label{eq:UV_luminosity}
\end{equation}
 where $f_\text{EUV}$ and $\alpha_{\text{EUV}}$
are the normalization constants and power-law indices of the SED approximated piecewise as a power law. Integrating this over frequency after dividing by photon energy gives us the photon emissivity that will be set by the calibration of the simulation, as discussed in Section~\ref{sec:calibrate},
 \begin{align}
  \dot{N}=\int_{\nu_{13.6}}^{\nu_{\infty}}\frac{L_\text{UV}}{h\nu}\mathrm{d}\nu.
  \label{eqn:findf}
 \end{align}  
 This allows us to get the value of the normalization $f_\text{EUV}$. As the UV luminosity is calculated at 1450\,\AA, we calculate $L_\text{UV}$ at this wavelength using a broken power law as described in \citet{Lusso2015}, with $\alpha_\text{EUV,1450}=-0.61$ for wavelengths $912$\AA\ $\lesssim \lambda \lesssim 1450$ \AA\ and $\alpha_\text{EUV,912}=-1.7$ for wavelength $\lesssim 912$ \AA. This gives us an intrinsic ionizing emissivity of the AGN in our simulation and is equal to the observed emissivity as we normalize to  an escape fraction of 1 for the AGN. The relation between  $M_\text{UV}$ and ionizing emissivity, assuming  canonical values of $\alpha_\text{EUV,1450}=-0.61$, $\alpha_\text{EUV,912}=-1.7$, and $f_\text{esc}=1$ is given by,
\begin{align}
    M_{\text{UV}} &= -16.31-2.5\log_{10}\left(\frac{\dot{N}}{10^{53}\text{photons/s}}\right)+2.5\log_{10}\left(\frac{f_\text{esc}}{1.0}\right)+\\\nonumber
    &0.503(\alpha_{(\text{EUV},1450)}+0.61)-2.5\,\text{log}_{10}\left(\frac{-\alpha_{(\text{EUV},912)}}{1.7}\right)
\end{align}
Lastly, to get the combined UV luminosity $M_\text{UV}$ of the faint AGN and the massive stars in their host galaxy, we add the AGN luminosity to the host galaxy luminosity before using Equation \ref{eq:UV_luminosity_2}. This is done independently for all AGN in the simulation.

What determines the ionizing emissivity for a given UV luminosity is the product  $f_{\rm esc}\times \xi_{\rm ion}$. 
We  varied$ f_\mathrm{esc}\xi_\mathrm{ion}$
until the luminosity function of the galaxies matched by eye the observed luminosity function. The results are shown in Figure~\ref{fig:uv_luminosity} 
(see \citet{Simmonds2024-2} for recent measurements of $f_\mathrm{esc}\xi_\mathrm{ion}$ at these redshifts).

We show the inferred UVLF in pink from our QSO-assisted simulations ($z\,=\,5.11,\,5.95,\,7.14,\,8.15,\,9.02,\,10.14$) with  
$f_\text{esc}\times\xi_\text{ion}\,=\,(3.07,\, 9.36,\,14.9,\,17.93,\,23.99,\,34.53)\times{10^{23}}$. The observed  UVLFs are taken from \citet{Oesch2018, Bouwens-2021, Donnan2024}. As discussed in \citet{Asthana2024C} for the ionizing emissivity in our simulation the inferred escape fractions required to match the observed galaxy UVLFs are rather low because of the observed rather high values of $\xi_{\rm ion}$ \citep{Simmonds2024-2}. In dotted pink, we also show the inferred UVLF for the AGN in the QSO-assisted model, and in the dashed pink curve, we display the UVLF of the AGN and massive stars combined for the galaxies that host AGN. The AGN data points are plotted as blue points and are taken from \citet{Maiolino2023b, Matthee2024, Harikane2023, Greene2024, Kokorev2024}. There are no observational data points for AGN at $z=9\text{ and }z=10$. Assuming an escape fraction of 1 for the AGN, our assumed UV luminosity functions of galaxies hosting AGN appear consistent with the admittedly still very uncertain observed AGN UVLFs at 1450\AA. Furthermore, at $z\sim 6$, the faint AGN in our simulations contribute less than 20\% of the total UV luminosity of the galaxies hosting them. This fraction decreases to less than 10\% at lower redshift and increases to 30\%  at $z=10$. Note that these values are due to our particular and somewhat arbitrary assumptions of how we include faint AGN in the simulations. The relatively low AGN contribution to UV luminosity of the galaxy hosting them in our QSO-assisted model also appears consistent with the -- we should stress again -- still very uncertain observations. 

In our QSO-assisted model, this low contribution is due to the assumed much larger escape fraction of photons emitted at 1450~\AA\ by accretion in AGN than by massive stars in galaxies. Furthermore, the effective $\xi_{\rm ion}$ of accretion inferred for our assumed SEDs of the AGN is about a factor three higher than the $\xi_{\rm ion}$  inferred from the observations assuming the ionizing emissivity is entirely due to massive stars \citep{Simmonds2024-2}. Note also that the UVLF of galaxies hosting AGN  spans a smaller magnitude range than the overall galaxy UVLF. This is because we assume the AGN to be hosted preferably in the most massive haloes in the simulation that host galaxies with higher than average luminosities. 

We show the AGN UVLF from the QSO-only model with the solid purple line. Because 10\% of galaxies host AGN in this model, the curve is higher than the number density of AGN in the QSO-assisted model (dashed pink line). Furthermore, this model has brighter and fainter AGN than the QSO-assisted model  as  expected. Overall, the distribution of AGN in the QSO-only model is also within the observational limits.

\subsection{The masses and accretion rates of the massive black holes powering the faint AGN}

The ionizing emissivity powered by accretion in the AGN can be related to the mass of the accreting black hole assuming a ratio of total luminosity of the AGN  to the Eddington luminosity, $f_{\rm Edd}$. Following \citet{mortlock2011} and \citet{willott2010}, we assume a bolometric correction of 4.4, and with our assumptions for the SED of the AGN in the UV  
this gives, 

\begin{equation} 
  \log_{10} M_{\text{bh}} = 7.27  + \log_{10}\left(\frac{\dot{N}}{10^{53}\text{photons s}^{-1}}\right)- \log_{10}\left(\frac{f_\text{edd}}{0.01}\right).
\end{equation}

The typical black hole masses inferred from JWST data for faint AGN, radiating at a rather small fraction of the Eddington limit, would thus be sufficient to produce the ionizing emissivity we have assumed. Note further that for Eddington-limited accretion with an efficiency of turning rest mass energy into radiation of ten percent, the e-folding or `Salpeter' time for the growth of a black hole is about 45 Myr. As assumed in our simulations, an ionizing emissivity due to AGN will thus not contribute significantly to the growth of the black holes discovered by JWST assumed to power the AGN. This is similar to what is inferred from the sizes of the near-zones around their brighter cousins that also suggest that the ionizing radiation these bright QSOs emit does not significantly contribute to the growth of the black holes powering them and that they have to mostly grow in an optical/UV obscured accretion mode\citep{eilers2017, satyavolu2023}.

\subsection{Constraining the contribution of faint AGN to reionization}

As mentioned above, the most robust constraint on QSO-assisted hydrogen reionisation is the expected early reionization of \heii\ to \heiii. This is due to the hard photons produced by accretion onto supermassive black holes. As discussed in detail by \citet{Madau2024}, the EUV SED of QSOs and the faint AGN discovered by JWST is very uncertain. As discussed by \citet{Lusso2015}, there are significant object-to-object variations in the EUV SED of QSOs. Furthermore nothing is known observationally about the EUV SED of fainter QSOs and AGN at higher redshifts. Note  that  \citet{Madau2024} assume that $f_{\rm esc}$ = 0.82 for hydrogen-ionizing photons and argue  that the fraction of \heiii\ ionizing photons escaping from the vicinity of the black holes and the host galaxy is also very uncertain. It is likely to be smaller than that for \hi\ ionizing photons because of the factor of about five larger recombination rate of \heiii\ compared to \hii. Panel D of Figure~\ref{fig:mosaic_quasar} shows the evolution of our models' \heiii\ volume filling factor and the AGN-only models of \citet{Madau2024}. Our QSO-assisted and QSO-only models reach 10\%  and 20\%, respectively, while the \citet{Madau2024} models reach about 30\%. \citet{Madau2024} argue that their models are not necessarily in conflict with \heiii\ reionization completing at $z\sim 2.7 -4$ as suggested by \heii\ opacity measurements and temperature measurements of the IGM \citep{Worseck2016, Laplante2018, Puchwein2019, Gaikwad2021, Becker2021, Makan2021, Makan2022}. Looking at the temperature evolution of our AGN models in panel~G of Figure~\ref{fig:mosaic_quasar} suggests, however, that the models with such large early \heiii\ fraction will not be consistent with the temperature measurements. 

Our QSO-assisted model with 17\% AGN contribution appears marginally consistent (see  \citeyearname{Puchwein2019} for a more detailed discussion of the temperature evolution of QSO-assisted and QSO-only models and \citeyearname{Basu2024} for a recent simulation of the later stages of helium reionization). For a larger AGN contribution to the ionizing emissivity to be consistent with the temperature measurements,  we would need a softer intrinsic SED and/or smaller escape fraction of \heii\ ionizing photons, similar to the findings of \citet{Madau2024}. This would, however,  have little effect on hydrogen reionization and not solve the problem of the mismatch of the width of the \hi\ \lya\ opacity distribution. Note also again that the optically thin approximation that we used for the photo-heating rates may somewhat underestimate the temperatures. 

Note also that \citet{Dayal2024} discussed a semi-numerical reionization model with 23\% contribution from faint AGN towards the end of reionization. Their model ends reionization somewhat too early to be consistent with the distribution of effective optical depth \citep[see also][]{Trebitsch2021, Trebitsch2023}

\citet{Chardin2017} presented a QSO-only model that reproduced earlier \hi\ \lya\ opacity measurements. However, they had to assume a much larger \hi\ ionizing emissivity dominated by brighter QSOs than the faint AGN in our simulations. They assumed furthermore a very short mean free path that was not self-consistently modelled. The \citet{Chardin2017} model would also not evade the temperature constraints without assuming a very soft EUV SED and/or small \heiii\ escape fraction.  Similar  results have been found by \citet{DAloisio2017}, \citet{Mitra2017} and \citet{Garaldi2019} for their QSO-only models. 

Note further that  additional constraints on the AGN contribution to the UV background could come from the rapid evolution of the incidence of CIV absorber at $4<z<6$ (see section 4.2.1 of Davies et al. 2024 for a discussion of the connection to ionizing \heii\ to \heiii\ by AGN.)

\subsection{Caveats}
Our  radiative transfer simulations  with \atmf\ are  performed by post-processing hydro-dynamical simulations performed with the homogeneous UV background from \citet{Puchwein-2023}. The inhomogeneous photo-heating during reionization and its effects on the thermal state of the IGM and the corresponding different hydrodynamic responses are not considered. As discussed in previous papers describing our \at\ simulations, the pressure smoothing scale of the IGM is relatively small, and the decoupling does not greatly influence the large-scale spatial distribution of neutral and ionized regions \citep{Kulkarni2015, Puchwein-2023}. Conversely, these simulations are significantly faster and 
enable careful calibration of the \lya\ forest data. \citep{Kulkarni-2019, Keating-2019}. \citet{Puchwein-2023} have proposed a hybrid approach where the hydrodynamic response of the IGM to inhomogeneous reionization is iteratively captured. This is done by employing hydro-simulations post-processed with \textsc{aton}, then re-running these simulations with spatially varying photo-heating rates derived from the post-processed outputs.

The AGN properties implemented in our models involved some rather simplistic choices, most notably fixing the AGN contribution to the ionizing emissivity to a fixed fraction independent of redshift,  fixing the duty cycle independent of halo mass, and choosing a fixed halo occupation fraction. While plausible first guesses, these could and should undoubtedly be varied in future work. As seen in the differences between QSO-assisted and QSO-only models, these assumptions do not only affect the thermal history of the IGM but also have a strong effect on the flux CDF. The spectral representation of the faint AGN should be also considered  a first best guess, anchored in what we know about much brighter QSO at lower redshift, which could and should be varied in future work.  

The choice representing the energy of the ionizing photons produced by  AGN and massive stars with just four frequency bins was necessitated by resource limitations and the calibration to the \lya\ forest simulation requiring several iterations of the simulations. Rerunning the simulations with more frequency bins to better handle the spectral hardening of ionizing UV background would be a worthwhile but resource-intensive project. Our QSO models should be seen as a pilot study.

The limited box size of the simulation restricts the inclusion of brighter QSOs, while the available resolution of our simulation results in predictions that are not fully converged, which in particular affects the accurate modelling of the mean-free path of ionizing photons \citep{Feron-2024}. 


\section{Conclusions} 
\label{sec:conclusion}

Motivated by the recent discovery by JWST of a large number density of accreting black holes during the epoch of reionization, we have included in our simulations with \atmf\ the effect of faint AGN. These are modelled as transient sources with a lifetime of 10~Myr, emitting ionizing radiation with spectra that are sufficiently hard to result in an early partial reionization of \heii\ to \heiii. We investigate a QSO-assisted model where we randomly chose 1\% from the massive haloes to host an AGN-powered ionizing source in addition to the source of ionizing radiation produced by the massive stars of its host galaxy. We further assume that the combined AGN ionizing emissivity is 17\% of the total ionizing emissivity from the stellar sources. We also study a QSO-only model where we chose 10\% of the massive haloes to place transient AGN, which emit all the ionizing photons. There is no contribution to the total ionizing emissivity from massive stars in the QSO-only model. The new QSO-assisted and QSO-only models have been calibrated to the mean \lya\ forest transmission at  $5\lesssim z\lesssim6.2$, similar to our previous galaxy-only models to which we compare.  Note further that our conclusions are for specific assumptions of the lifetime of the AGN. Establishing the robustness of these results will require exploration of a wide parameter space which is beyond the scope of our work here \citep{satyavolu2023}. Our main findings are as follows.


\begin{itemize}
\item  The QSO-only model requires a lower emissivity by a factor $1.8$ than the galaxy-only models towards the end of reionization and, unlike in the galaxy-only models it requires a rather shallow drop of the ionizing emissivity as reionization completes. The QSO-assisted model delays the completion of hydrogen reionization to $z\sim5$ with more neutral islands remaining until $z\sim5.4$,  compared to the galaxy-only models where reionization ends at $z\sim5.2$. The QSO-only model does not complete reionization until $z=5$ and has neutral islands even at $z=5$. Our QSO-assisted model provides a good fit to the observed distribution of the \lya\ optical depths. On the other hand, our QSO-only model appears to be completely inconsistent with the width of the \lya\ optical depth distribution and future work is required to establish whether this is generally the case for QSO-only models.  

\item The number of \heiii\ ionizing photons is about six times larger in the QSO-only model compared to the QSO-assisted model.  Accordingly, the volume-weighted \heiii\ fraction at $z\sim6$ is 5\% in the QSO-only model and 30\% in the QSO-assisted model. Additionally, in the QSO-only model, where galaxies do not contribute to reionization, the scarcity of \hei\ ionizing photons creates a situation where, despite the increase in \heiii\ ionizing photons, the lack of \hei\ ionization hinders the expansion of \heiii\ regions. Note that the transient nature of the AGN results in the \heiii\ regions growing much more inhomogeneously than the \hii\ regions. 

\item Both AGN models have a higher IGM temperature than the galaxy-only models primarily due to an early onset of \heii\ reionization, with a small effect from ionization of hydrogen by harder photons. The temperature in the QSO-assisted model is within the observational constraints and starts to decrease at $z\sim5.4$, compared to the galaxy models in which the decrease starts at $z\sim5.6$. However, the QSO-only model results in too high and still increasing IGM temperatures at $z\lesssim 5$. Note that the very inhomogeneous nature of \heiii\ reionization thereby leads to increased spatial fluctuations of the temperature-density relation of the IGM. Note further that the exact value of the temperature depends on the assumed SED of the AGN.

\item Given the observed ionizing efficiencies $\xi_\text{ion} $, the galaxies in our QSO-assisted models require a rather low  escape fraction to match the \lya\ forest data. For a given UV luminosity AGN can thus be a factor of up to fifty times more efficient in producing hydrogen-ionizing photons if the escape fraction of Lyman continuum photons from AGN approaches unity. A contribution by AGN of 20\% of the galaxy ionizing emissivity in 1\% of  JWST-detected reionization-epoch galaxies, as seen in our QSO-assisted model, appears in reasonable agreement with what we have learned from JWST about faint AGN at high redshift so far. A contribution at this level does not contribute significantly to the growth of the central supermassive black holes in the AGN host galaxies. Note, that the escape fractions are not self-consistently modelled but instead are chosen to match the unfortunately still rather uncertain UVLF inferred from observed faint AGN.
\end{itemize}

Our results suggest that a modest contribution to reionization by faint AGN, as indicated by JWST observations, is in good agreement with the \lya\ forest data. In contrast, reionization dominated by faint AGN appears to be challenging to reconcile with the \lya\ optical depths distribution during the late stages of reionization. The QSO-only model is likely also inconsistent with the observed temperature measurements of the IGM at lower redshift. 

\section*{Acknowledgements}
We thank Christopher Cain, Nick Gnedin, Vid Ir\v{s}i\v{c}, Piero Madau, Brant Robertson and Sindhu Satyavolu for helpful discussions. The work was performed partially using the Cambridge Service for Data Driven Discovery (CSD3), part of which is operated by the University of Cambridge Research Computing on behalf of the STFC DiRAC HPC Facility (\href{www.dirac.ac.uk}{www.dirac.ac.uk}). The project was also supported by a Swiss National Supercomputing Centre (CSCS) grant under project ID s1114. This research was supported in part by grant NSF PHY-2309135 to the Kavli Institute for Theoretical Physics (KITP). Support by ERC Advanced Grant 320596 ‘The Emergence of Structure During the Epoch of Reionization’ is gratefully acknowledged. MGH has been supported by STFC consolidated grants ST/N000927/1 and ST/S000623/1. GK gratefully acknowledges support from the Max Planck Society via a partner group grant. GK is also partly supported by the Department of Atomic Energy (Government of India) research project with Project Identification Number RTI 4002. The work has been performed as part of the DAE-STFC collaboration 'Building Indo-UK collaborations towards the Square Kilometre Array' (STFC grant reference ST/Y004191/1). SA also thanks the Science and Technology Facilities Council for a PhD studentship (STFC grant reference ST/W507362/1) and the University of Cambridge for providing a UKRI International Fees Bursary. This research used resources of the Oak Ridge Leadership Computing Facility at the Oak Ridge National Laboratory, which is supported by the Office of Science of the U.S. Department of Energy under Contract No. DE-AC05-00OR22725. These resources were granted via INCITE AST206.

\section*{Data availability}
All data and analysis code used in this work are available from the first author upon request.  

\bibliographystyle{mnras}
\bibliography{refs} 

\appendix
\section{ Zooming in on a moderately bright AGN.}
\begin{figure*}
  \centering
  \includegraphics[page=1,width=\linewidth,trim={9cm 10cm 8cm 10cm},clip]{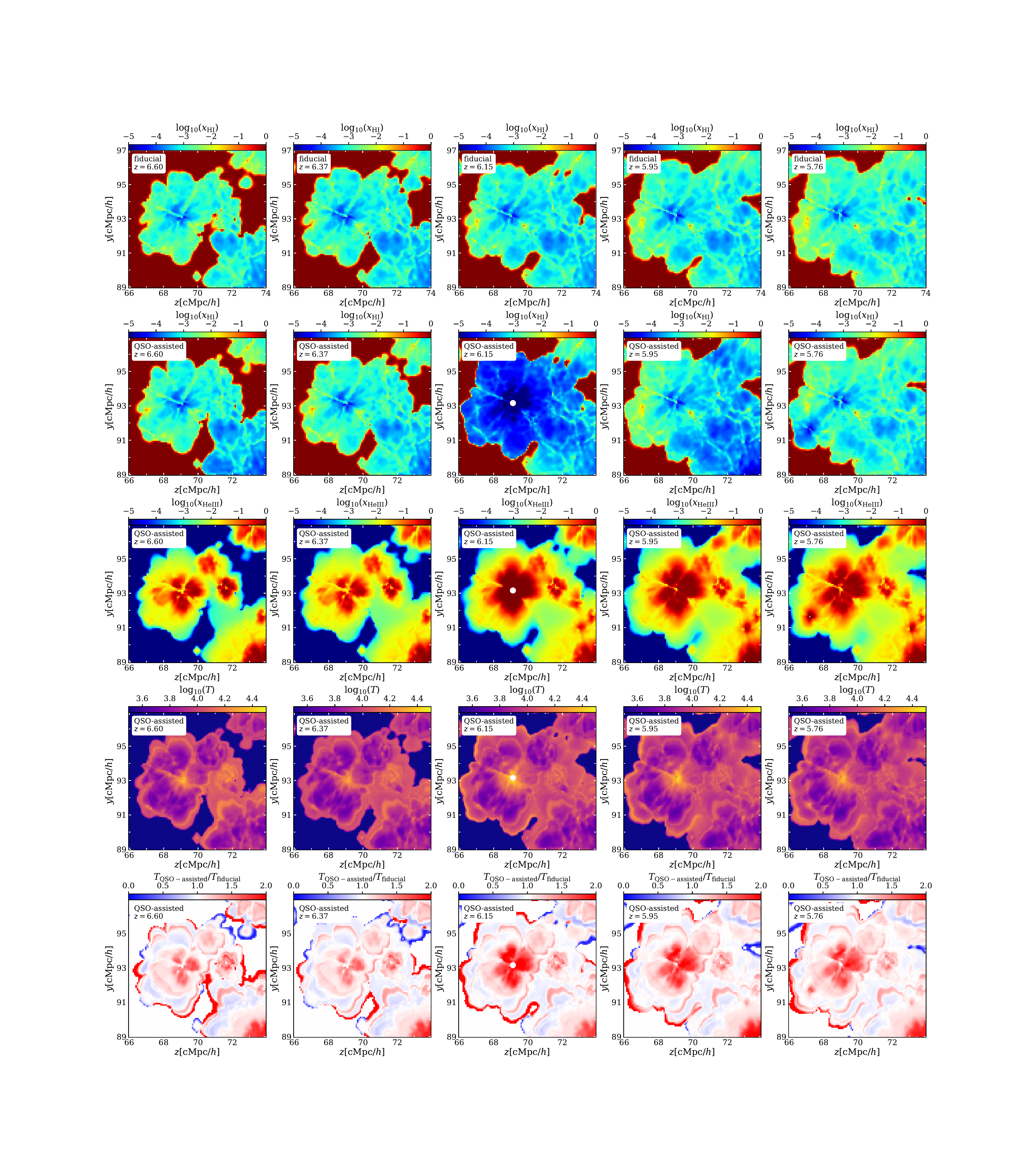}
  \caption{The evolution of neutral hydrogen, \heiii\ and temperature around in a small part of the simulation box where a QSO turns on in the QSO-assisted model at $z\,=\,6.20$ and emits until $z\,=\,6.15$. The AGN has an absolute magnitude $M_\text{UV,1450}=-18.74$, and is emitting $9.34\times10^{53}\text{ ionizing\, photons\, s}^{-1}$. The top two rows show the evolution of neutral hydrogen in the fiducial model (no AGN) and in the QSO-assisted model. The middle row shows the evolution of the \heiii\ fraction, while the bottom two rows show the temperature for the QSO-assisted and the ratio of the temperature to that in the fiducial model in ionized cells with temperatures greater than $10^{3.4}$~K.  The white dot marks the AGN at the centre of the zoom. The five columns show the same location at $z\,=\,6.60,\,6.36,\,6.15,\,5.95\text{, and }5.76$. }
  \label{fig:zoom_qso}
\end{figure*}

In Figure~\ref{fig:zoom_qso}, we show a zoom in-region around a part of the simulation box hosting a moderately bright AGN in the QSO-assisted model, turning on at $z\,=\,6.20$ and stopping to  emit at $z\,=\,6.15$. The $\dot{N_{\rm ion}}$ is $9.34\times10^{53}\text{ photons s}^{-1}$, and the $M_{1450}=-18.74$. The top and bottom rows show the neutral hydrogen fraction and temperature in the fiducial model. The middle three rows show the neutral fraction, \heiii\ fraction and the temperature at the same location for the QS0-assisted model. When the AGN turns on, a clearly visible region develops around the AGN where hydrogen and \heii\ temporally become more highly ionized, and the injection of hard photons from emission by the AGN leads to additional photo-heating. The temperature of the fiducial model is shown as the ratio to that in the QSO-assisted model. The different effect on \hi\ and \heiii\ because of the five times larger recombination rate of the latter is clearly visible. 

\section{The masses of the  DM haloes hosting AGN}\label{sec:mass_distribution}
\begin{figure*}
    \centering
    \includegraphics[width=1.5\columnwidth,page=1]{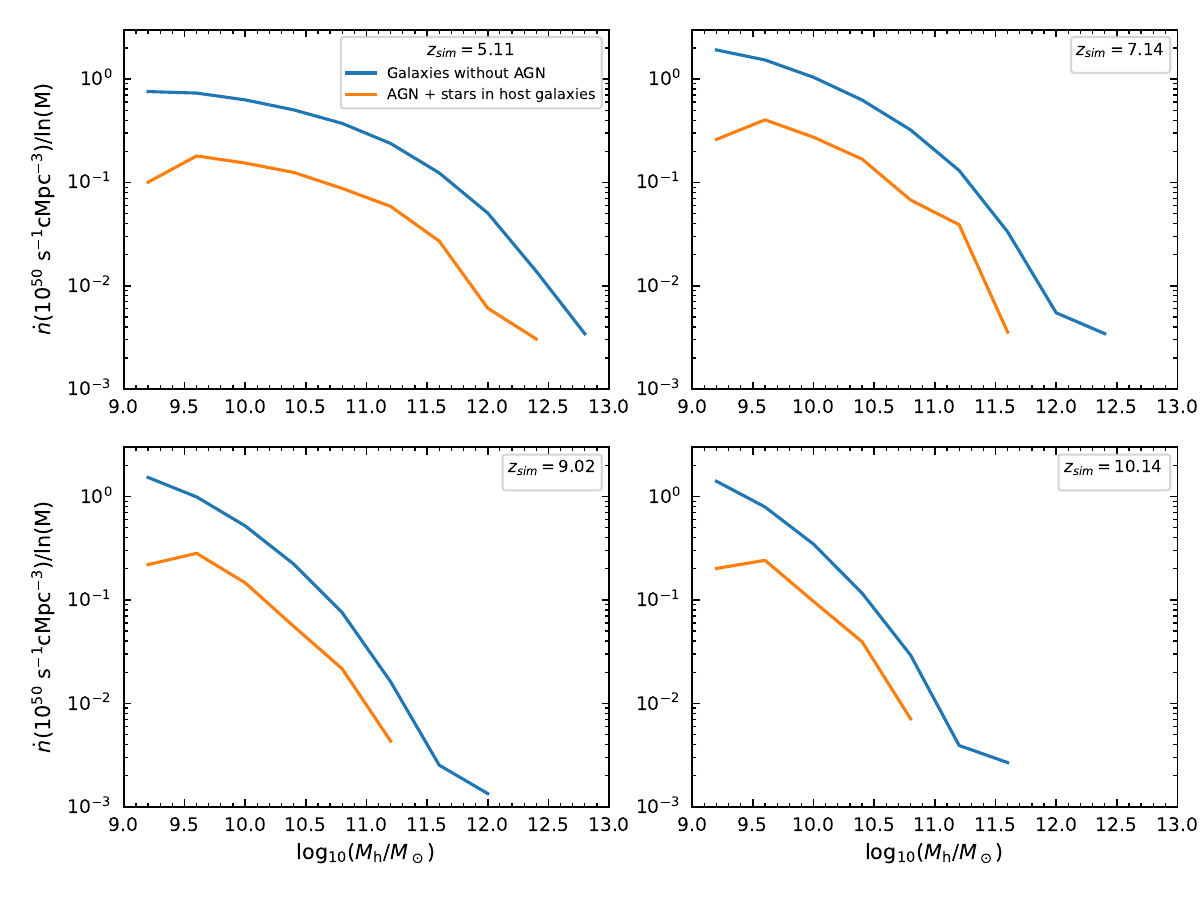}
    \caption{The emissivity as a function of halo mass for the QSO-assisted model at four different redshifts. The blue curves show the contribution from galaxies, while the orange curves show the contribution from AGN plus stars  in their host galaxies.}
    \label{fig:mass_distribution}
\end{figure*}
The maximum mass of haloes that exist within the simulation changes from $\sim 10^{10}~M_\odot$ at $z\sim13$ to $\sim 10^{13}~M_\odot$ at $z\sim5$ (see Figure 1 of \citeyearname{Asthana2024} ). Thus, the mass range of haloes that host QSOs also changes with redshift. Efficient growth of supermassive black holes requires the presence of a deep potential well. It is therefore generally assumed that the AGN prefer more massive haloes. To account for this, we have placed the AGN in the haloes, starting with the most massive haloes and populating haloes of lower mass until they add up to our chosen AGN fraction. The simulation achieves this by first calculating the number of haloes between  $10^{12}-10^{13} M_\odot$. If the number of haloes in this range is insufficient, the lower limit is decreased by a factor of 2 and the process is repeated. The condition of a sufficient number of haloes is decided based on the fraction of haloes hosting an AGN. For our QSO-assisted model, we choose 1\% of the haloes to host an AGN with a lifetime of 10~Myr. As the snapshots are updated every 40~Myr, we choose a mass range to contain at least $n=1\times40/10$\% of the total sources. However, this condition is also insufficient. As the simulation proceeds, the most massive haloes will accrete mass and thus will mostly remain the most massive haloes. As a result, the same sources will be hosting an AGN as the simulation proceeds. To prevent this, we multiply $n$ with an arbitrary constant (10 in our model), so the number of haloes from which to choose a halo that will host an AGN  is much larger. In Figure \ref{fig:mass_distribution}, we show the contribution of the volume emissivity emitted by haloes in a given mass range for four different redshifts. The blue curves represent the galaxies in the QSO-assisted model, while the orange curves represent the AGN plus stars in their host galaxies. The majority of hydrogen ionizing photons emitted by massive stars are coming from the lowest mass galaxies in the simulation. Focusing on the orange curves, we can see that in all four panels, there is a  drop in the emissivity contribution from low-mass haloes. We also see that the most massive halo hosting a QSO changes from $10^{12.5}~M_\odot$ at $z=5.11$ to $10^{10.5}~M_\odot$ at $z=10.14$. This is expected as the maximum halo mass increases with decreasing redshift. Lastly, the probability of selecting a halo  to host an AGN is low due to the small number of the most massive haloes. Thus, the most massive halo to host an AGN  is not the same as the most massive halo that emits ionizing photons. 

\section{Differences in the PDF  of the neutral fraction}\label{appen:pdf}

\begin{figure*}
    \centering
\includegraphics[width=\linewidth,page=1]{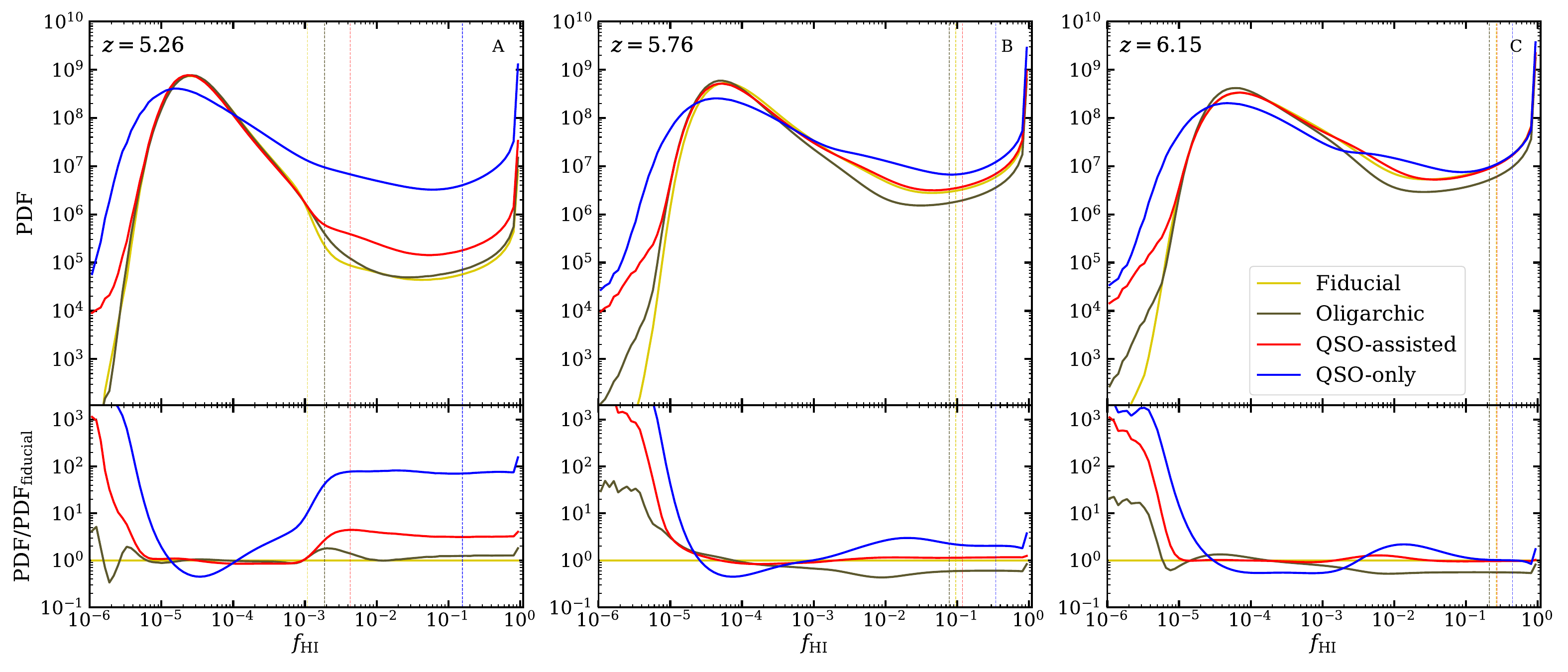}
    \caption{\textit{Top:} the probability distribution function of the neutral fraction in the four simulation models at three different redshifts. \textit{Bottom:} the ratio between the PDF of the models to the fiducial model. The vertical lines show the volume-averaged neutral fraction for each model.}
    \label{fig:pdf_xhi}
\end{figure*}

In Figure~\ref{fig:mosaic_quasar}, we had seen  that even so the mean \lya\ transmission of the models has been calibrated to have the same value, the volume-averaged neutral fractions of the QSO-assisted and QSO-only can be up to  two and three orders of magnitude different from those of the 
galaxy-only fiducial and Oligarchic model. To give further insight into this  we show in Figure~\ref{fig:pdf_xhi} the probability distribution function of the neutral fraction for the four different models: fiducial, Oligarchic, QSO-assisted and QSO-only in yellow, brown, red and blue respectively. The distribution is shown at three different redshifts to illustrate the evolution as reionization finishes. As expected,  in the QSO-only and QSO-assisted models the number of neutral cells is disappearing more slowly and the two models with AGN have a tail that extends to lower neutral fraction than  in the galaxy-only models. As discussed in section 4 the top panel of  Fig.6 also shows subtle but clearly visible differences in the spatial distribution of the neutral fraction. 

\section{Conversion to ionizing emissivity}\label{appen:madau_cal}

In panel C of Figure~\ref{fig:mosaic_quasar}, we show the ionizing emissivity from \citet{Madau2024}. The conversion to ionizing emissivity injected into the IGM ($\text{photons s}^{-1}\text{Mpc}^{-3}$), from the comoving emissivity ($\text{erg s}^{-1}\text{Hz}^{-1}\text{Mpc}^{-3}$) provided in their paper is given by,
\begin{align}
  \dot{n} = \bar{f}_\text{esc,H}\int_{\nu_{912}}^{\nu_{228}} \frac{\epsilon_{\nu}}{h\nu}\,d\nu\, +\, \bar{f}_\text{esc,He}\int_{\nu_{228}}^{\nu_\text{np.infty}} \frac{\epsilon_{\nu}}{h\nu}\,d\nu,
  \label{eqn:ionizing_emissivity}
\end{align}
where $\dot{n}$ is the total ionizing emissivity, $\epsilon_{\nu}$ is the intrinsic comoving emissivity, $\bar{f}$ are the escape fractions for hydrogen and helium, and $\nu_i$ and $\nu_f$ are the lower and upper limits of the integral. The lower limit of the first integral is the Lyman limit for hydrogen, and the upper limit is the helium ionizing frequency. The second integral extends from the helium ionizing frequency to infinity. The intrinsic comoving emissivity is given by, 
\begin{equation}
  \epsilon_{\nu}\, =\, (1-\bar{f}_\text{host})\,\epsilon_{912}\,(\nu/\nu_{912})^{\alpha_\text{EUV}},
  \label{eqn:comoving_emissivity}
\end{equation}
where $\epsilon_{912}$ is the emissivity at 912\AA, $\alpha_\text{EUV}$ is the spectral index, and $\bar{f}_\text{host}$ is the population-averaged fractional input of the host galaxy to the rest-frame UV light.Putting Equation~\ref{eqn:comoving_emissivity} into Equation~\ref{eqn:ionizing_emissivity} we get,
\begin{align}
  \dot{n}= &\frac{(1-\bar{f}_\text{host})\,\epsilon_{912}}{h}\,\left(\bar{f}_\text{esc,H}\,\int_{\nu_{912}}^{\nu_{228}} \frac{(\nu/\nu_{912})^{\alpha}}{\nu}\,d\nu \right)+ \\\nonumber
  &\frac{(1-\bar{f}_\text{host})\,\epsilon_{912}}{h}\,\left(\bar{f}_\text{esc,He}\int_{\nu_{228}}^{\nu_\text{np.infty}} \frac{(\nu/\nu_{912})^{\alpha}}{\nu}\,d\nu,\right).
\end{align}
Integrating the equation, and knowing $\alpha$ is a negative number, we get,
\begin{equation}
  \dot{n} = \frac{(1-\bar{f}_\text{host})\,\epsilon_{912}}{h\alpha}\left(\bar{f}_\text{esc,H} \left(\left(\frac{912}{228}\right)^{\alpha} - 1 \right) - \bar{f}_\text{esc,He}  \left(\frac{912}{228}\right)^{\alpha} \right).
\end{equation}
For the two models, the $\bar{f}_\text{host}$ is set to 0.4, while the escape fractions for hydrogen and helium are taken to be 0.8 and 0.3, respectively, with the spectral index $\alpha$ set to -1.4. In the second model, the escape fraction is 0.9 for both, and the spectral index is -1.9.

\section{Speed up of RT using GPU streams}
\at\ is a highly parallelized GPU based code originally described in \citet{Aubert-2008, Aubert-2010}. GPUs have since evolved, and we have optimized the code for the current generation of NVIDIA GPUs. This was achieved by enabling non-blocking communication between processes and incorporating the concept of pinned memory, asynchronous memory transfers and GPU streams in \atmf. Non-blocking communication is functions (\texttt{MPI\_Isend, MPI\_Irecv}) that allow programs to initiate a send or receive operation and proceed with other work before the communication completes. A stream in CUDA is a sequence of operations that execute in order on the GPU. By default, all operations are placed in the default stream, which executes them sequentially. However, CUDA supports multiple streams that can execute concurrently on the GPU. Asynchronous memory copy (\texttt{cudaMemcpyAsync}) is a function that copies data between the host and device memory asynchronously. This means the CPU does not wait for the memory copy to be completed before proceeding to the next step. Lastly, one can allocate pinned host memory by using \texttt{cudaMallocHost}. This type of memory is not paged out by the operating system and is required for asynchronous memory copy. Implementing these techniques allowed us to achieve a speed-up of 30\%, which is discussed below.

To solve the radiative transfer (RT) equations in post-processing, the simulation volume is broken into sub-cubes on which the RT is solved on GPUs. At the end of every time step,  information about photons on the boundary of the sub-cubes is communicated to the neighbouring sub-cubes. Looking at the process involved on a single sub-cube, the following steps are involved: we first have to copy the data from GPUs to CPUs, then we have to communicate the data to the neighbouring sub-cubes, and finally, we have to copy the data back to the GPUs that we have received from the neighbouring sub-cubes. These steps have to be repeated for every neighbour of the sub-cube, which can be up to 6 neighbours in 3D. 

In the original code implementation, the memory copies between the host and device are synchronous. This means the CPU waits for the memory copy to be completed before proceeding to the next step. Communication between neighbouring sub-cubes is also blocked, which means the CPU waits for the communication to be completed before proceeding to the next step. Lastly, all the above steps are repeated for the next neighbour but only after the previous neighbour has been completed. In \atmf, every single step mentioned above is parallelized. By using streams and asynchronous memory copy, we are simultaneously copying data from the GPUs to CPUs (and vice-versa) for all the neighbours simultaneously. By implementing non-blocking communication, the copied data is communicated to the neighbouring sub-cubes while the CPU is doing other work.

\bsp	
\label{lastpage}

\end{document}